\documentclass[10pt, journal, compsoc]{IEEEtran}

\ifCLASSINFOpdf
  \usepackage[pdftex]{graphicx}
  \DeclareGraphicsExtensions{.pdf,.jpeg,.png}
\else
  \usepackage[dvips]{graphicx}
  \DeclareGraphicsExtensions{.eps}
\fi

\ifCLASSOPTIONcompsoc
  \usepackage[nocompress]{cite}
\else
  \usepackage{cite}
\fi

\usepackage{hyperref}
\usepackage{algpseudocode}
\usepackage{algorithm}
\usepackage[font=normalsize,labelfont=sf,textfont=sf]{caption}
\usepackage[font=footnotesize,labelfont=sf,textfont=sf]{subcaption}
\usepackage{multirow}
\usepackage{tabularx}
\usepackage{hhline}
\usepackage{amsmath}
\usepackage{epstopdf}
\usepackage{xcolor}

\hypersetup{pdfauthor={Author's name},%
pdftitle={Document Title},%
pagebackref=true,%
colorlinks=true,%
pdftex
}

\makeatletter
\renewcommand{\ALG@name}{Protocol}
\makeatother

\begin{document}

\title{TeamPhone: Networking Smartphones for Disaster Recovery}

\author{


Zongqing~Lu, Guohong~Cao, and~Thomas~La~Porta

\IEEEcompsocitemizethanks{
\IEEEcompsocthanksitem  Z. Lu, G. Cao and T. La Porta are with the Department of Computer Science and Engineering, The Pennsylvania State University, University Park, PA 16802. E-mail: \{zongqing, gcao, tlp\}@cse.psu.edu.}}

\IEEEcompsoctitleabstractindextext{
\begin{abstract}

In this paper, we investigate how to network smartphones for providing communications in disaster recovery. By bridging the gaps among different kinds of wireless networks, we have designed and implemented a system called TeamPhone, which provides smartphones the capabilities of communications in disaster recovery. Specifically, TeamPhone consists of two components: a messaging system and a self-rescue system. The messaging system integrates cellular networking, ad-hoc networking and opportunistic networking seamlessly, and enables communications among rescue workers. The self-rescue system groups, schedules and positions the smartphones of trapped survivors. Such a group of smartphones can cooperatively wake up and send out emergency messages in an energy-efficient manner with their location and position information so as to assist rescue operations. We have implemented TeamPhone as a prototype application on the Android platform and deployed it on off-the-shelf smartphones. Experimental results demonstrate that TeamPhone can properly fulfill communication requirements and greatly facilitate rescue operations in disaster recovery.
\end{abstract}

}

\maketitle
\IEEEdisplaynotcompsoctitleabstractindextext
\IEEEpeerreviewmaketitle

\ifCLASSOPTIONcompsoc
\IEEEraisesectionheading{\section{Introduction}\label{sec:intr}}
\else
\section{Introduction}
\label{sec:intr}
\fi

\IEEEPARstart{D}{uring} last decade, many communication technologies have been applied to improve rescue efforts following a disaster, such as deploying wireless sensor networks for emergency response \cite{lorincz2004sensor,george2010distressnet} and employing smart badges to form a mobile ad-hoc network and then gathering information from trapped survivors of structural collapse \cite{zussman2003energy}. However, as learned from the 2011 Great East Japan earthquake, the only helpful services in disaster recovery are those that are used daily by everyone \cite{fujihara2014disaster}. To provide communications in disaster recovery, smartphones, equipped with both cellular and short-range radios (\emph{e.g.}, WiFi, Bluetooth), are the most promising communication tools. Although cellular towers might also be destroyed by disasters, {\em e.g.}, in the 2008 Sichuan earthquake \cite{sichuan2008}, short-range radios of smartphones can still provide communications. Moreover, the ubiquity of smartphones further opens great opportunities to reinvestigate disaster recovery from the network point of view.

In disaster recovery, smartphones have the potential to be the most feasible communication tools. For example, trapped survivors of a structural collapse can communicate with rescue workers and report their position information through the short-range radio (\emph{e.g.,} WiFi) of their smartphones when they are within the communication range of each other. Smartphones of rescue workers can also form networks using WiFi and meet the communication needs in disaster recovery. 

To this end, in this paper, we propose TeamPhone, a platform for communications in disaster recovery, where smartphones are teamed up and work together to provide data communications. By exploiting WiFi and cellular modules of smartphones, TeamPhone seamlessly integrates cellular networking, ad-hoc networking and opportunistic networking, and supports data communications among rescue workers in infrastructure-constrained and infrastructure-less scenarios. TeamPhone also enables energy-efficient methods for trapped survivors to discover rescue workers and send out emergency messages, by carefully addressing the wake-up scheduling of smartphones. The emergency message includes the coarse-grained location and position information of trapped survivors, which is derived from the last known locations of their smartphones and the network formed by these smartphones. We implement TeamPhone as an app on the Android platform and deploy it on off-the-shelf smartphones. Experimental results demonstrate that TeamPhone can properly fulfill the communication requirements and greatly facilitate rescue operations.

The main contribution of this paper is the design, implementation and evaluation of TeamPhone. This contribution breaks down into the following aspects:
\begin{itemize}
\vspace{-0.1cm}
\item We design TeamPhone which consists of a messaging system and a self-rescue system to provide communications and facilitate rescue operations in disaster recovery. The idea of TeamPhone is motivated by the fact that people heavily reply on smartphones in their daily lives. 

\item The messaging system can accomplish different types of message transmissions by bridging cellular networks, ad hoc networks and opportunistic networks, and by connecting different routing protocols.

\item The self-rescue system can send out emergency messages with location and position information through self-rescue grouping, wake-up scheduling and positioning, where we design a communications protocol that can fulfill these functions in an energy-efficient manner.

\item The design, implementation and evaluation are based on off-the-shelf smartphones, which enables TeamPhone to be installed as a factory default application on smartphones by manufacturers to facilitate rescue operations in disaster scenarios. 
\end{itemize}

%
%

The rest of this paper is structured as follows. We start with the motivation and challenges in Section \ref{sec:moti}, and then present the design and implementation of TeamPhone in Section \ref{sec:desi} and Section \ref{sec:impl}, respectively. Then, we present experimental evaluations in Section \ref{sec:eval}, followed by the review of related work in Section \ref{sec:rela} and the conclusion in Section \ref{sec:conc}. 

\section{Motivation and Challenges}
\label{sec:moti}
In this section, we first motivate the need of networking smartphones in disaster recovery for communications, and then illustrate the challenges.

\subsection{Motivation}
Disasters, such as earthquakes, may topple countless homes and kill thousands of people. Power failures and fallen cellular towers caused by disasters further leave the affected area cut off from the outside and hinder rescue operations. In disaster recovery, communications are crucial for coordinating rescue operations. Moreover, if trapped survivors in the rubble can send out emergency messages to rescue workers, rescue operations can be greatly accelerated. Therefore, in this paper, we investigate how to provide communications in disaster recovery.

With the increasing penetration of smartphones equipped with short-range radios, GPS and sensors, smartphones have been studied for various applications, including health monitoring \cite{sun2015symdetector}, mobile sensing \cite{ra2012medusa}, ad-hoc communications \cite{shi2012serendipity}, WiFi-based localization \cite{wang2014eyes}, \emph{etc}. Moreover, smartphones have evolved to be much more powerful than before in terms of computing and communications, and users heavily rely on smartphones in their daily lives. As a result, users always carry their smartphones or place smartphones where they can easily and immediately be accessed even during disasters. However, in disaster recovery such as earthquakes, the cellular towers may be destroyed, and thus cellular communication of smartphones is cut off. Then, we have to set up communication with short-range radios (\emph{e.g.}, WiFi) of smartphones. 

Smartphones have recently been conceptually considered for disaster recovery to locate immobilized survivors using Bluetooth in \cite{suzuki2012soscast} and to provide multi-hop communications in \cite{nishiyama2014relay}. However, Bluetooth has limited communication range (a few meters). The design in \cite{suzuki2012soscast} fails to consider energy efficiency, which may quickly drain the battery of the smartphone. \cite{nishiyama2014relay} also fails to conserve energy and uses proactive routing which reacts slowly to the frequently changing network topology in disaster recovery. It also increases the maintenance overhead in terms of network traffic and energy consumption. In contrast to these existing works, we propose a much more functional and energy-efficient communication system using smartphones for disaster recovery, and we address various design and implementation issues. 


\subsection{Challenges}
\label{sec:challenges}
During disaster recovery, communications satellites and mobile cellular towers may be deployed. However, communications satellites are scarce. 
Although mobile cellular towers can be used to set up the command center and provide critical communications for rescue workers, such 
vehicle carried towers cannot cover all disaster areas. Therefore, it is important to network smartphones with short range radios 
in disaster recovery.

Due to the mobility of rescue crews and survivors, network topology changes frequently; \emph{i.e.}, sometimes smartphones may form a mobile ad-hoc network, and sometimes they only contact each other opportunistically. Therefore, the big challenge is how to provide communication crossing different types of networks including ad-hoc networks, opportunistic networks, and cellular networks, considering frequent topology changes; \emph{i.e.}, rescue workers can communicate with each other and with the command center no matter if they are within the coverage of the mobile cellular towers or not.

In addition, trapped survivors may be buried in debris and be difficult to discover. With the capability of communications between smartphones, the devices of trapped survivors can automatically send out emergency messages to nearby rescue crews with better discoverability and reachability. However, broadcasting emergency messages can drain batteries quickly. Since rescue operations may last for days after disasters occur, the discovery of nearby rescue crews and sending out emergency messages in an energy-efficient manner is another challenge. 


Moreover, it is better for trapped survivors to include their location information into emergency messages to facilitate rescue operations. However, in disaster recovery, GPS and network providers are not available for localization. Providing some coarse-regained location or position information of trapped survivors, which can yet be exploited for rescue operations, is also challenging.


\begin{figure*}[!t]
\setlength{\belowcaptionskip}{-17pt}
\setlength{\abovecaptionskip}{5pt}
\begin{minipage}[t]{.38\textwidth}
	\centering
    	\includegraphics[width=1\textwidth]{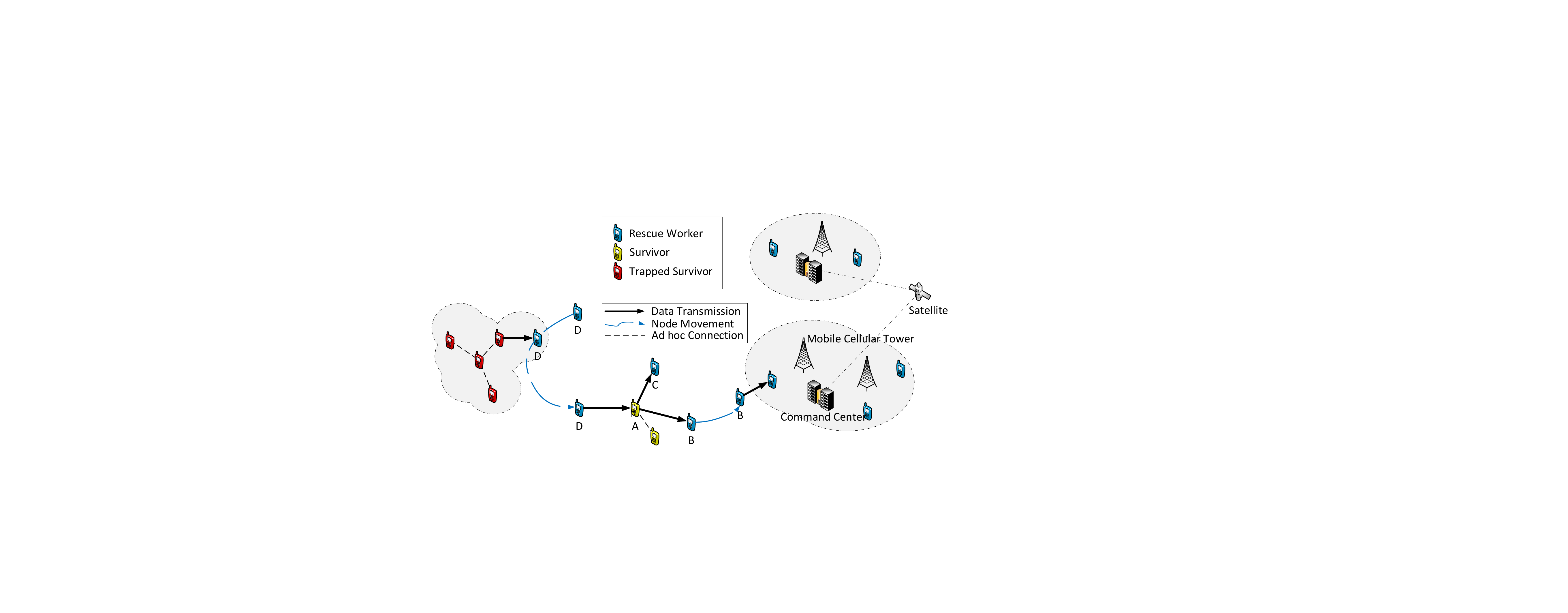}
    	\caption{Network scenario in disaster recovery}
    	\label{fig:scenario}
\end{minipage}
\hspace{.1cm}
\begin{minipage}[t]{.60\textwidth}
	\centering
	\includegraphics[width=1\textwidth]{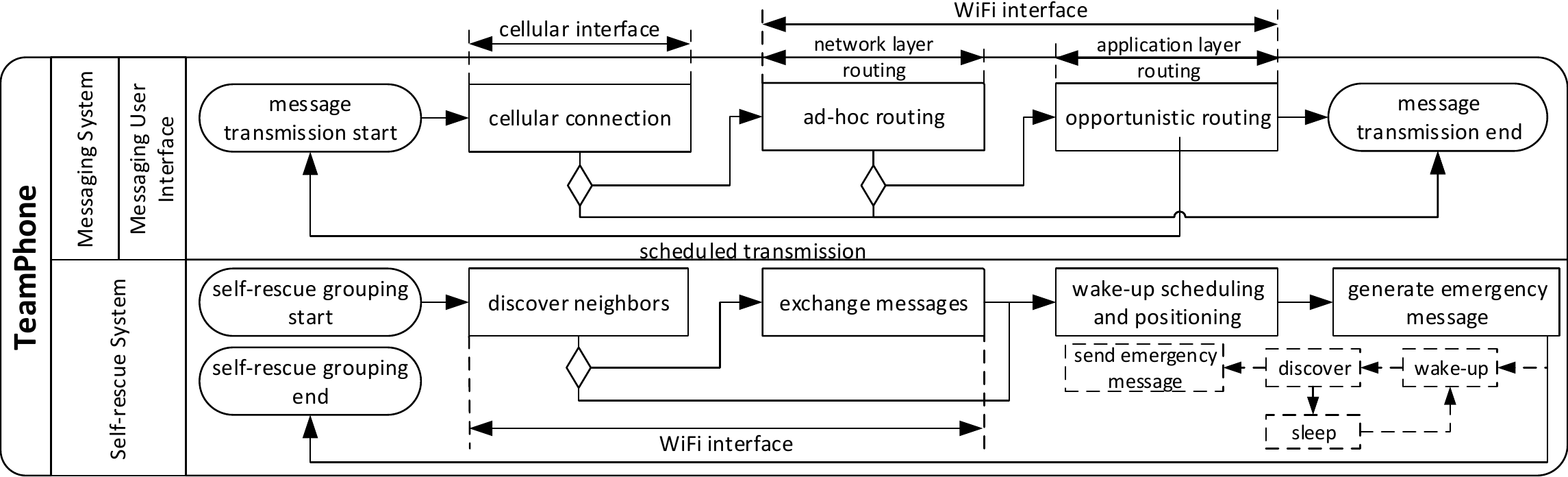}
    	\caption{Overview of TeamPhone}
    	\label{fig:architecture}
\end{minipage}
\end{figure*}

\section{TeamPhone}
\label{sec:desi}
TeamPhone is a tailored system for disaster recovery based on smartphones, which provides seamless data communication through several different types of networks and facilitates rescue operations for trapped survivors. In this section, we describe the network scenario of disaster recovery, and present the design of TeamPhone.

\subsection{Network Scenario}

\figurename~\ref{fig:scenario} illustrates the network scenario for disaster recovery. As can be seen, mobile cellular towers only cover limited areas and provide cellular communications for rescue workers in these areas. A command center sits in the covered region and command centers in different regions may be connected via communications satellites. Survivors can also join the network to help data communications. Trapped survivors in the debris are unable to move and are waiting for rescue.

When rescue workers fall out of cellular coverage, they can only use short-range radios (\emph{e.g.}, WiFi) to communicate. They can communicate with each other or with the command center via individuals or combinations of ad-hoc connections, opportunistic contacts, and cellular connections, as shown in \figurename~\ref{fig:scenario}. Also, trapped survivors can construct a group based on ad-hoc connections and send out emergency messages when rescue workers or survivors are within the communication range of the group. Therefore, the exploitation of smartphones can greatly extend the communication field far beyond the region covered by mobile cellular towers, and increase the opportunity of being discovered and rescued for trapped survivors.

\subsection{Overview}
As shown in \figurename~\ref{fig:architecture}, TeamPhone includes two components: the messaging system and the self-rescue system. The messaging system runs on the smartphones of rescue workers or survivors and provides message transmissions. The self-rescue system runs on the smartphones of trapped survivors, which automatically forms groups with nearby trapped survivors, performs positioning and sends out emergency messages.

TeamPhone works as follows. First, users need to specify which system to use. If smartphones are specified as part of the messaging system (we call them messaging nodes), users can send messages via their smartphones. Their smartphones will act as relays for both ad-hoc routing and opportunistic routing, and as gateways when they have cellular connections. The smartphones of trapped survivors will be automatically configured as part of the self-rescue system (we call them self-rescue nodes), which is triggered by other apps that measure seismic waves, such as \textit{iShake} \cite{iShake}. Self-rescue nodes are grouped and scheduled to wake up and receive \textit{hello} messages from messaging nodes. When self-rescue nodes receive \textit{hello} messages, they will automatically generate an emergency message and send it to the messaging node, and then the messaging node can send the emergency message to the command center. Messaging nodes can also initiate emergency messages if needed.  

\subsection{Messaging System}

The messaging system is designed to handle message transmissions via three ways: through cellular connections, by ad-hoc communications and upon opportunistic contacts. When a messaging node needs to transmit a message (text, voice, photo, \emph{etc.}), it first tries to reach the destination via the cellular network. The message can be delivered only when both the source and destination are within the region covered by mobile cellular towers. 
If direct transmission to the cellular network fails (e.g., in the case the sender is out of the cellular coverage), the messaging system will try to reach the destination by the ad-hoc network and through the cellular network via ad hoc relays, \emph{i.e.}, the messaging node will issue a routing request to construct a routing path to the destination based ad-hoc communications and cellular connections. If the request is fulfilled, the message can be directly sent to the destination. 

Since messaging nodes that have cellular connections act as gateways in the ad-hoc network, they can be exploited to reach the destinations that have cellular connections, such as the command center. Therefore, the message might be transmitted to the gateway by ad-hoc connections and then relayed by the gateway to the destination by cellular connections. If there is no routing path to the destination but the command center can be connected, the message will be sent to and stored at the command center and it will be forwarded to the destination once it enters the cellular region. Otherwise, the message will be stored locally and transmitted upon opportunistic contacts between messaging nodes, and different opportunistic routing strategies can be applied. 

The IP addresses of rescue workers and survivors for their WiFi interfaces can be assigned by the mobile cellular tower when they connect with the cellular network. For example, when a device first connects to the cellular network, the mobile cellular tower will send out a static IP address for its WiFi interface, and the information of other devices (e.g., the smartphones of rescue workers) that have been already assigned an IP address is shared to the device. The information is updated each time the device directly connects to the cellular network. In this way, survivors and rescue workers are aware of each other. Moreover, we do not assume that any traffic goes beyond the cellular network and the network formed by WiFi of smartphones. The traffic between these two networks is handled by gateways as discussed above. Therefore, NAT and Internet public IPs are not needed.


\subsubsection{Ad-hoc Routing}
\label{sec:aodv}
To reach the command center via ad-hoc connections, messages must be relayed at a gateway, \emph{i.e.}, a messaging node in the cellular region. Suppose a reactive routing protocol is employed (\emph{e.g.}, AODV routing). The messaging nodes will reply to the routing request differently, based on whether they are within or outside of the cellular region. Due to node movement, messaging nodes need to monitor the status of their cellular connection and configure themselves as gateways for the routing protocol when they have cellular connections, or vice versa.

In the following, we use AODV as an example to illustrate how to bridge the ad-hoc network and cellular network via ad-hoc routing in our design. When a gateway receives a routing request for a destination, if it does not have an active path to the destination, in addition to forwarding the routing request to its neighbors, it will also send back a routing reply with a `gateway' flag and the hop count to the source. If the source receives a routing reply with an ad-hoc path to the destination before the timeout, it will ignore the routing replies with the `gateway' flag. Otherwise, the source will select the gateway with the minimum hop count, encapsulate the message and send it to the gateway. When the gateway receives the message, it will decapsulate it, connect the destination and forward the message through its cellular interface.

\subsubsection{Opportunistic Routing}
In the messaging system, opportunistic routing acts as an alternative when the destination cannot be connected via cellular or ad-hoc communications. Opportunistic routing works as an application that forwards messages between two nodes that encounter each other. The opportunistic routing of the messaging system adopts two simple forwarding strategies: \textit{(i)} static routing where the message carried by a messaging node is forwarded only when it encounters the destination, to save network resources such as energy and bandwidth; \textit{(ii)} flood routing (also known as epidemic routing) where messaging nodes that carry the message always forward it to the encountered node such that the delay of the message can be minimized. More sophisticated routing schemes tailored for rescue operation are to be considered in future work. Note that TeamPhone is a system that can adopt any opportunistic routing protocol.

Unlike mobile opportunistic networks where nodes are isolated individually, the network in disaster recovery most likely consists of groups of nodes (\emph{e.g.}, groups of rescue workers), where nodes within the group are well connected via ad-hoc communications and nodes between groups can also be connected through other nodes, \emph{e.g.}, smartphones of survivors. In such a scenario, ad-hoc routing is preferred for message transmission rather than opportunistic routing. Therefore, in the messaging system, opportunistic routing is explored only when network partitions occur and thus two simple routing strategies are adopted rather than other sophisticated schemes based on historical contact information, such as \cite{lu2014skeleton}\cite{lu2015algorithms}.

\subsubsection{Routing Efforts}
\label{sec:efforts}
The messaging system differentiates messages into two categories: general messages and emergency messages (by using a flag in the message), on which the messaging system makes different routing efforts. Currently, the system treats messages from self-rescue nodes as emergency messages and all other messages as general messages. General messages are handled considering the balance between network resource and delay, whereas emergency messages are deliberately handled to minimize the delay. Different strategies are adopted to handle these two types of messages.

General messages are handled in the order of cellular connections, ad-hoc communications and opportunistic contacts as aforementioned. If a general message has to be forwarded opportunistically, then static routing is adopted.

Emergency messages that are initiated by trapped survivors will be first received at nearby messaging nodes and then forwarded to the command center for better coordination of rescue operations. 
Since trapped survivors are usually far from the regions covered by cellular towers, cellular is usually unavailable for nearby messaging nodes to send out the emergency message. If the command center also cannot be connected by ad-hoc communications, the emergency message will be handled as follows. The nearby messaging node will forward a copy of the emergency message to all the messaging nodes that can currently be connected via ad-hoc communications. Then, at each node the emergency message is stored and forwarded via opportunistic contacts by the flood routing. Whenever a node receives an emergency message, it will first examine whether it can forward the emergency message to the command center directly by a cellular or ad-hoc path. If not, it will continue the flood routing. 


\subsection{Self-rescue System}
The smartphones of trapped survivors are configured as part of the self-rescue system and then the self-rescue system automatically sends out emergency messages when rescue workers or survivors are nearby. The battery life of smartphones must last as long as possible, since rescue operations may last for hours or even days. Therefore, the self-rescue system must be energy-efficient. Since trapped survivors are most likely difficult to discover, rescue crews may not infer the location of trapped survivors, even if they have received emergency messages from them. Thus, the emergency message should also provide location information to facilitate rescue operations.

\subsubsection{Self-rescue Grouping and Wake-up Scheduling}
\label{sec:schedule}
To save energy, instead of continuously staying awake, a self-rescue node can wake up periodically to discover messaging nodes. However, this will increase the possibility that a self-rescue node is asleep when a messaging node passes by. Since survivors in the same building may be trapped together or nearby when the building collapses, they can group together as a {\em self-rescue group} via WiFi and wake up in a coordinated way to save energy. 

The group coordination must be carefully done, since they may still miss passing by nodes if only one node in the group is scheduled to wake up during some time. For example, as shown in \figurename~\ref{fig:scenario}, node $D$ moves around the group of trapped survivors. If the leftmost node is scheduled to wake up and all others are sleeping, the self-rescue node will not receive the \textit{hello} message since it is out of the communication range of node $D$. Therefore, the design of the self-rescue system should take this into consideration. Note that our wake-up scheduling problem is similar to the sensor activity scheduling problem \cite{wang2011coverage} in sensor networks. However, existing work focuses on achieving full coverage by scheduling among redundant nodes, satisfying coverage ratio requirement, and maximizing coverage lifetime. Unlike these problems, in disaster scenarios, a self-rescue group most likely does not have redundant nodes (i.e., impossible to achieve full coverage by scheduling), and there is no imposed requirements of coverage ratio and lifetime. Given a network randomly formed by smartphones, our goal is achieve a good tradeoff between coverage and energy cost, considering the characteristics of disaster scenarios. In the following we present the details of our solution.


\textbf{Clique.} Considering a group of nodes that coordinate waking up, we need to decide which nodes should be awake during each time period so as to reduce energy cost while maintaining good coverage. To do so, we consider cliques. In a clique there exits an edge between every two nodes, and hence any node in a clique can cover all other nodes. Therefore, in a clique nodes are close to each other, and the area covered by one node takes a large proportion of the area covered by all nodes in the clique.

\begin{figure}[t]
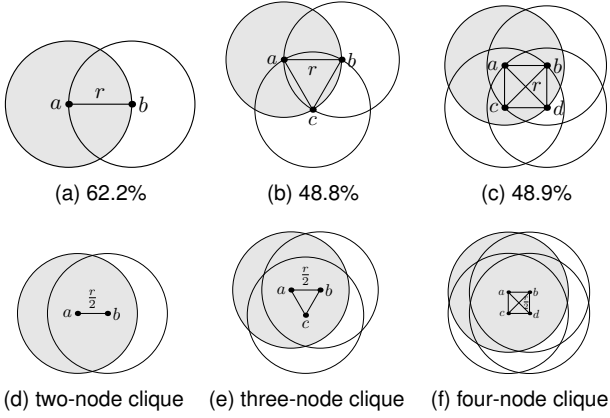

\setlength{\abovecaptionskip}{3pt}
\centering
	\begin{subfigure}[t]{.14\textwidth}
		\centering
    	\includegraphics[width=1\textwidth]{TeamPhone/TeamPhone.3}
    	\caption{62.2\%}
    	\label{fig:two-node}
    \end{subfigure}
    \begin{subfigure}[t]{0.16\textwidth}
		\centering    	
    	\includegraphics[width=.8\textwidth]{TeamPhone/TeamPhone.4}
    	\caption{48.8\%}
    	\label{fig:three-node}
    \end{subfigure}
    \hspace{.1cm}
    \begin{subfigure}[t]{0.12\textwidth}
		\centering    	
    	\includegraphics[width=1\textwidth]{TeamPhone/TeamPhone.5}
    	\caption{ 48.9\%}
    	\label{fig:four-node}
    \end{subfigure}
	
	\vspace{0.3cm}	
	
	\begin{subfigure}[t]{.14\textwidth}
		\centering
    	\includegraphics[width=.8\textwidth]{TeamPhone/TeamPhone.7}
    	\caption{two-node clique}
    	\label{fig:two-node-r/2}
    \end{subfigure}
    \hspace{0.05cm}
    \begin{subfigure}[t]{0.145\textwidth}
		\centering    	
    	\includegraphics[width=.75\textwidth]{TeamPhone/TeamPhone.8}
    	\caption{three-node clique}
    	\label{fig:three-node-r/2}
    \end{subfigure}
    \hspace{0.05cm}
    \begin{subfigure}[t]{0.14\textwidth}
		\centering    	
    	\includegraphics[width=.75\textwidth]{TeamPhone/TeamPhone.9}
    	\caption{four-node clique}
    	\label{fig:four-node-r/2}
    \end{subfigure}
\caption{Illustration of wireless coverage of one node in two-node clique, three-node clique and four-node clique, where $r$ is the transmission radius.}
\label{fig:clique}
\end{figure}

\figurename~\ref{fig:clique} illustrates the ratios between the coverage of one node and the coverage of all nodes in two-node, three-node and four-node cliques. In Figures~\ref{fig:two-node}, \ref{fig:three-node} and \ref{fig:four-node}, the percentage indicates the least coverage ratio for each case; \emph{i.e.} the least coverage ratio for two-node clique, three-node clique and four-node clique are 62.2\%, 48.8\% and 48.9\%, respectively, when the longest distance between nodes is exactly the transmission radius $r$. The coverage ratio increases with the decrease of such distance. For example, when it is decreased to $r/2$, as shown in Figures~\ref{fig:two-node-r/2}, \ref{fig:three-node-r/2} and \ref{fig:four-node-r/2}, the coverage ratio of one node greatly increases and it is close to the coverage of all nodes in the clique. Thus, instead of waking up individually, nodes in a clique should wake up alternately (\emph{i.e.}, there is no more than one awake node at any time in a clique) to discover nearby messaging nodes so as to save energy.

\begin{figure*}[!t]
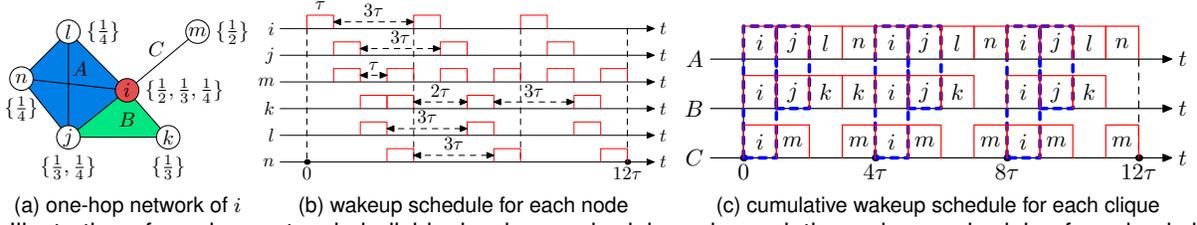

\setlength{\abovecaptionskip}{15pt}
\setlength{\belowcaptionskip}{-15pt}
\centering
	\begin{subfigure}[t]{.18\textwidth}
		\centering
    	\includegraphics[width=1\textwidth]{TeamPhone/TeamPhone.0}
    	\caption{one-hop network of $i$}
    	\label{fig:onehop network}
    \end{subfigure}
    \begin{subfigure}[t]{0.30\textwidth}
		\centering    	
    	\includegraphics[width=1\textwidth]{TeamPhone/TeamPhone.1}
    	\caption{wakeup schedule for each node}
    	\label{fig:onehop wakeup}
    \end{subfigure}
    \hspace{0.025cm}
    \begin{subfigure}[t]{0.37\textwidth}
		\centering    	
    	\includegraphics[width=1\textwidth]{TeamPhone/TeamPhone.2}
    	\caption{cumulative wakeup schedule for each clique}
    	\label{fig:cumulative wakeup}
    \end{subfigure}
\caption{Illustration of one-hop network, individual wakeup schedule and cumulative wakeup schedule of maximal cliques.}
\label{fig:onehop}
\end{figure*}

\textbf{Wake-up among Maximal Independent Sets.} Since a network can be seen as being composed of maximal cliques\footnote{ A maximal clique is a clique to which no more nodes can be added to form another clique.}, {\em e.g.}, the network shown in \figurename~\ref{fig:onehop network} consists of three maximal cliques which are $A=\{i,j,l,n\}$, $B=\{i,j,k\}$ and $C=\{i,m\}$. The problem is how to choose a node from each maximal clique to wake up together so as to yield better coverage. Using \figurename~\ref{fig:onehop network} as an example, let us consider two node sets $\{n, k, m\}$ and $\{j, k, m\}$, where three nodes of each node set are from these three maximal cliques, respectively. Comparing these two node sets, $\{n, k, m\}$ is better than $\{j, k, m\}$ since the coverage of node $k$ overlaps more with the coverage of node $j$ than node $n$. In other words, it is because nodes $j$ and $k$ are adjacent. Therefore, it is better to select the nodes from the maximal cliques such that they are not adjacent. Such a node set is also called the \emph{maximal independent set}\footnote{A maximal independent set is an independent set such that adding any other node to the set forces the set to contain two adjacent nodes.}.  For the network in \figurename~\ref{fig:onehop network}, the maximal independent sets are $\{i\}$, $\{j,m\}$, $\{l,k,m\}$ and $\{n,k,m\}$, whose union equals to the node set of the network. A maximal independent set with more nodes has larger coverage. However, to balance the energy consumption at each node, the maximal independent sets have to be scheduled to wake up alternately.

To determine the wake-up schedule for a network, we need to find all maximal independent sets. However, finding all maximal independent sets is a NP-hard problem and requires global information of the entire network. Moreover, as we will discuss later, wake-up scheduling among maximal independent sets may not save much energy for some nodes, compared to being always awake. To this end, we break this NP-hard problem into easier problems and determine the wake-up scheduling at each node in a distributed way. More specifically, each node builds a one-hop network, finds all maximal cliques, and determines the wake-up schedule based the schedules of other nodes within a maximal clique. In the following, we first briefly describe how to build the one-hop network at each node.

\textbf{One-hop Network.} Self-rescue nodes broadcast beacon messages to learn their neighbors. As the configuration of self-rescue nodes is triggered automatically, e.g., by the seismic detection, self-rescue nodes most likely start the process simultaneously. After a short period time a node should have received beacon messages from all its one-hop neighbors. Let $N_u$ denote the set of one-hop neighbors of node $u$. Then, they broadcast their one-hop neighbor set such that all nodes know their two-hop neighbors. Note that self-rescue nodes do not consider unidirectional links; \emph{i.e.}, for example, if node $u$ receives the one-hop neighbor set of node $v$ but $N_v$ does not include $u$, node $u$ will not count node $v$ as a one-hop neighbor. Based on the information of two-hop neighbors, they can construct one-hop networks that include one-hop neighbors and edges among them, for example, as shown in \figurename~\ref{fig:onehop network}, which is built by node $i$. Note that to determine the maximal cliques a node belongs to, the one-hop network is sufficient, since nodes in a clique must be fully connected.

\textbf{Scheduling within/across Maximal Cliques.} Each node acknowledges its one-hop network and it can also find all maximal cliques in the one-hop network. Although finding all maximal cliques is also NP-hard (it is complement to finding all maximal independent sets) \cite{eppstein2013listing}, it can be easily performed by a node since the one-hop network is usually small regardless of the size of the rescue group. In a maximal independent set, there are no two adjacent nodes; \emph{i.e.}, there has to be no more than one awake node at any time in a maximal clique. Therefore, our scheduling should follow this rule.

Moreover, when all maximal independent sets are scheduled alternately, the wake-ups may not be well balanced among nodes since some nodes may belong to multiple maximal independent sets. For example, for the network shown in \figurename~\ref{fig:onehop network}, as discussed above there are four maximal independent sets (\emph{i.e.}, $\{i\}$, $\{j,m\}$, $\{l,k,m\}$ and $\{n,k,m\}$) and node $m$ belongs to three of them, and hence the wake-up frequency of node $m$ is $\frac{3}{4}$ (\emph{i.e.}, $\frac{3}{4}$ of total rotations). This only saves $\frac{1}{4}$ energy for node $m$ compared to always staying awake. Therefore, we have to consider a way to further save energy for such nodes.

A node may belong to multiple maximal cliques and thus the node should determine a wake-up schedule across all these cliques. For example, in \figurename~\ref{fig:onehop network}, node $i$ belongs to cliques $A$, $B$ and $C$ and the wake-up fraction of node $i$ is $1/4$ (since there are four nodes in clique $A$), $1/3$ and $1/2$ for $A$, $B$ and $C$, respectively. To save energy, for each node, we choose the lowest wake-up fraction of the maximal cliques it belongs to, denoted as $\gamma$, as the wake-up frequency (\emph{e.g.}, $\gamma_i=1/4$ for node $i$). By doing so, the wake-up frequency for all nodes is less than or equal to the frequency using maximal independent sets. For example, the wake-up frequencies of nodes $m$ and $k$ are reduced from $\frac{3}{4}$ to $\frac{1}{2}$ and from $\frac{1}{2}$ to $\frac{1}{3}$, respectively.

Furthermore, since the wake-up schedules of self-rescue nodes in the same maximal clique are dependent, to determine the wake-up schedule at each node in a distributed way, we have to decide a scheduling order. We use the sum of wake-up fractions of the maximal cliques that a node belongs to as the criterion, denoted as $\theta$, to determine the scheduling order.  Within a maximal clique, the node with a larger $\theta$ decides first.


\figurename~\ref{fig:onehop wakeup} shows the wake-up schedule of each node in the network of \figurename~\ref{fig:onehop network}, assuming the wake-up schedule starts at time $0$ and each wake-up time period is $\tau$, which is a system parameter. The wake-up schedule of the nodes is easily determined following the scheduling order discussed above and two rules: \textit{(i)} the wake-up frequency for each node must be $\gamma$; and \textit{(ii)} there is no more than one wake-up node at any time in a clique. \figurename~\ref{fig:cumulative wakeup} shows the wake-up schedule for each clique, which is the cumulative schedule of nodes in each clique. We can see for cliques $B$ and $C$, there are vacancies because node $i$ is scheduled at $\gamma_i=1/4$ rather than $1/2$ in clique $C$ and node $j$ is scheduled at $\gamma_i=1/4$ rather than $1/3$ in clique $B$. Since the vacancy ratio is low, \emph{i.e.}, 16.7\% for clique $B$ and 25\% for clique $C$, and the coverage ratios are high as discussed above, the schedule vacancy only slightly affects the overall coverage; yet such wake-up scheduling can greatly save energy. 

Nodes may belong to more than one clique, such as nodes $i$ and $j$ in \figurename~\ref{fig:cumulative wakeup}. When such nodes are awake, no other nodes in its clique need to wake up to save energy. As an example, for the network of Figure~\ref{fig:onehop network}, the total number of scheduled wake-up periods of all the nodes is $22$ ($27$ if using maximal independent sets) for a time period $12\tau$ as shown in Figure \ref{fig:onehop wakeup}, while if all the nodes stay awake, the total number of wake-up periods is $72$. Therefore, we can see the scheduling can save about 70\% of the energy, assuming each node spends the same amount of energy for each wake-up period $\tau$; in other words, it can extend the battery life of the self-rescue group more than three times.

\textbf{Optimality} Obviously, keeping all nodes awake can guarantee the optimality of coverage, but drain the battery quickly, while keeping only one node awake can guarantee the optimality of efficiency (coverage/energy), i.e., no coverage overlaps, but with the minimal coverage. Assume self-rescue nodes have the same WiFi range and energy cost per wake-up period. As at most one node in a maximal clique is scheduled to wake up, there is no coverage overlap in the clique, but there may be overlaps with nodes from other cliques. 
The combination of awake nodes in the self-rescue group at a time determined by our solution maximizes the coverage with its efficiency, i.e., adding one more awake node cannot increase the coverage without decreasing the efficiency. The proof is straightforward. As each awake node is selected from an individual clique, additional node will contribute a larger overlap to the coverage and thus reduce the efficiency.  

\textbf{Distributed Wake-up Scheduling.} In the following, we describe how to determine the wake-up schedule in a distributed way. Note that the distributed scheduling works on the entire network, not just the one-hop network. 

As discussed above, by exchanging neighboring information, each node can construct the one-hop network and find all maximal cliques, and it can further calculate $\theta$ and $\gamma$ based on the maximal cliques. Then each node floods $\theta$ into the network, and after that each node acknowledges $\theta$ of all other nodes. The node that has the maximum $\theta$ will initiate the scheduling procedure; \emph{i.e.}, it decides a reasonable start time for the wake-up schedule, determines its own wake-up schedule based on $\gamma$, and then broadcasts the schedule to its one-hop neighbors. 

\begin{figure*}[!t]
\setlength{\abovecaptionskip}{15pt}
\setlength{\belowcaptionskip}{-15pt}
\centering
\begin{minipage}[t]{.64\textwidth}
\centering
	\begin{subfigure}[t]{.3\textwidth}
		\centering
    	\includegraphics[width=1\textwidth]{TeamPhone/TeamPhone.13}
    	\caption{one-hop network of $i$ and built coordinate system}
    	\label{fig:coordinate-1}
    \end{subfigure}
    \hspace{0.1cm}
    \begin{subfigure}[t]{.26\textwidth}
		\centering
    	\includegraphics[width=1\textwidth]{TeamPhone/TeamPhone.14}
    	\caption{one-hop network of $j$}
    	\label{fig:coordinate-2}
    \end{subfigure}
        \hspace{0.1cm}
    \begin{subfigure}[t]{.3\textwidth}
		\centering
    	\includegraphics[width=1\textwidth]{TeamPhone/TeamPhone.15}
    	\caption{positioned nodes in self-rescue group}
    	\label{fig:coordinate-3}
    \end{subfigure}   
\caption{Illustration of building the coordinate system for self-rescue group}
\label{fig:coordintate}
\end{minipage}
\hspace{0.05cm}
\begin{minipage}[t]{.30\textwidth}
\setlength{\abovecaptionskip}{5pt}
\centering
\includegraphics[width=.83\textwidth]{TeamPhone/TeamPhone.16}
\caption{Example of integrated wake-up scheduling and positioning in a self-rescue group}
\label{fig:example}
\end{minipage}
\end{figure*} 

The nodes that receive the schedule and belong to a same maximal clique, \emph{e.g.} $j$ and $k$ in clique $B$ in \figurename~\ref{fig:onehop network}, need to determine the scheduling order. Since $\theta_j > \theta_k$, node $j$ goes first. As node $k$ also acknowledges $\theta_j > \theta_k$, it will not proceed until it receives the schedule of $j$. For the nodes from different maximal cliques, \emph{e.g.} nodes $j$ and $m$, their schedules are independent and thus the scheduling order does not apply to them. Since node $i$ decides the start time of the wake-up schedule, clock synchronization is required for all the nodes in the self-rescue group. However, smartphones usually get the local time from network providers such that self-rescue nodes are already synchronized with millisecond accuracy before disasters occur. Therefore, no additional synchronization is required since $\tau$ is at the second level.

In summary, when a node receives a schedule from a neighboring node, it first checks whether it has received the schedules from all other nodes with larger $\theta$ in the same maximal clique of the sender. If it has, it will determine its own wake-up schedule based on the received schedules and and broadcast the schedule to its neighbors, otherwise it will wait for other schedules. If there are more than one node in the maximal clique that have the same value of $\theta$, the order is determined based on node \textit{id}. Eventually, each node has its wake-up schedule. 



\subsubsection{Location and Positioning}
\label{sec:position}

\textbf{Last Known Location.} When sending out emergency messages, it is desired to include location information of trapped survivors to better help rescue crews to find them. However, since trapped survivors are buried in the debris, GPS is not available. In addition, cellular towers are destroyed and thus localization by network providers \cite{rauscher2001wireless} is also unavailable. Therefore, for each individual smartphone, the only location information it has is the last known location which is kept by the operating system (\emph{e.g.}, the last known location that can be retrieved from the Android system consists of location, timestamp, accuracy and provider). Although many apps on smartphones require a user's location to be periodically updated in background, the last known location does not necessarily indicate current location. However, given a self-rescue group, there are still chances that the last known location is close to the current location for some self-rescue nodes. While self-rescue nodes themselves cannot determine that based on the last known location they currently have, it is easy for rescue workers to verify the location since they acknowledge their current location. Therefore, it is necessary to include all last known locations of the self-rescue group into emergency messages. 

\textbf{Coordinate System.} Moreover, based on the communications among self-rescue nodes, it is possible to build a relative coordinate system in which the positions of self-rescue nodes can be computed. Node positions together with last known locations can be very helpful for rescue crews to locate the trapped survivors quickly, \emph{e.g.}, the last known location close to current location can be plugged into the coordinate system to reveal the locations of other self-rescue nodes. In the following, we detail how to build the coordinate system for a self-rescue group.

Given a wireless network and the distances between directly connected nodes (one-hop neighbors), the coordinate system for the network can be built as investigated in \cite{capkun2001gps}. The distance between one-hop neighbors can be measured by different methods, such as Angle of Arrival, Time of Arrival and Time Difference of Arrival. However, due to the limitation of the WiFi chips in off-the-shelf smartphones, the only method that can currently be exploited is the Signal Strength method, but this method can be easily replaced whenever a more accurate method becomes available for off-the-shelf smartphones. The positioning method in \cite{capkun2001gps} is designed for mobile ad hoc networks (distances between nodes are required to be periodically updated and positions have to be re-computed periodically) but does not address communication overhead, which is indeed very important for the self-rescue system due to the energy constraint. Thus, we design a more efficient positioning scheme to establish a coordinate system for the self-rescue group. The positioning scheme is specifically designed on top of self-rescue grouping and scheduling so as to reduce communication overhead.  

First, when a node broadcasts beacon messages to build one-hop networks, it will also include the information (\emph{e.g.}, transmit power, antenna gain, etc.) that is required to compute the distance at receivers. When a node broadcasts its set of one-hop neighbors, it will also enclose the distances to these neighbors. Thus, a node will acknowledge the distance of each edge in its one-hop network. For example, in Fig~\ref{fig:onehop network}, node $i$ will know the distance of each edge in the network. Then, node $i$ chooses the maximal clique that contains the maximum number of its one-hop neighbors, \emph{i.e.} clique $A$, to define a coordinate system, as illustrated in \figurename~\ref{fig:coordinate-1}, and then all the nodes contained in clique $A$ can be positioned in the coordinate system by triangulation since all the distances are known (refer \cite{capkun2001gps} for detail). 

For other maximal cliques in the one-hop network, if they share an edge with the maximal clique that has been positioned, then they can be also positioned in the coordinate system, \emph{e.g.}, clique $B$ in \figurename~\ref{fig:coordinate-1}, though there are two possible positions for some nodes due to symmetry. For example, based on the coordinates of nodes $i$ and $j$ and the distances between $i$ and $k$ and between $j$ and $k$, another possible position of node $k$ is $k'$. However, if node $k$ is located at $k'$, it should have discovered nearby nodes $l$ and $n$, considering their distances in the coordinate system. Although it is possible in rare cases that node $k$ is at the position $k'$, node $k$ is most likely at the position shown in the figure. Therefore, the positioning scheme is designed to choose the most likely position if such cases occur. For the maximal clique that does not share an edge with the positioned clique, the position cannot be determined, \emph{e.g.}, node $m$ in clique $C$, which is called outlier. 

After computing the coordinate of each one-hop neighbor, node $i$ will send the coordinate information to its neighbors. The information also includes the distances to outliers, though they cannot be positioned by node $i$ in the coordinate system. However, outliers may be positioned by other nodes in the self-rescue group with the help of the included distance information. 

When node $j$ receives the information, it acknowledges its position in the coordinate system built by node $i$ and will use the information to compute the position of a node that is not yet positioned in its one-hop network, \emph{i.e.}, node $o$ as depicted in \figurename~\ref{fig:coordinate-2}. Finally, in the self-rescue group, only node $m$ is not positioned as illustrated in \figurename~\ref{fig:coordinate-3}. Although it is possible that some nodes in the self-rescue group cannot be positioned, the positioning scheme can provide the position information of self-rescue nodes as much as possible based on their connections.

\textbf{Self-rescue Group Communications Protocol.} In the following, we detail how to integrate the positioning scheme with the distributed wake-up scheduling so as to minimize the communication overhead. The positioning is designed to be initiated at the node with the maximum sum of wake-up fractions as the scheduling does. The node determines a coordinate system, computes the positions of its one-hop neighbors and broadcasts them together with the determined wake-up schedule to its neighbors. For all other nodes in the network, when it is ready to determine its own wake-up schedule, it will also compute the positions of its one-hop neighbors if needed (it is possible these nodes are already positioned). If the wake-up scheduling is completed, which means all of its one-hop neighbors are already scheduled, it will flood the position information it has into the network. Otherwise, it broadcasts the schedule and positions, and the process continues. Finally, all nodes acknowledge the position information of the self-rescue group.


Let us use \figurename~\ref{fig:example} as an example to illustrate how the wake-up scheduling and positioning work together in a self-rescue group, where the order of the process is labeled by red numbers. \textit{1)} Since the sum of wake-up fractions of node $i$ is the maximum, node $i$ will initially determine its wake-up schedule, setup a coordinate system, compute positions for its one-hop neighbors, and then broadcast the schedule and the position information, denoted by $(i^*,j^*,k^*,l^*,n^*,m)$, where $*$ indicates the position of the node is known in the coordinate system. \textit{2)} Then, nodes $m$ and $j$ will proceed to scheduling according to the rules discussed in Section \ref{sec:schedule}. Node $j$ will compute the position of node $o$ and then broadcast $(i^*,j^*,k^*,l^*,n^*,m, o^*)$. Since all nodes in the maximal clique node $m$ belongs to have scheduled wake-up, node $m$ will flood $(i^*,j^*,k^*,l^*,n^*,m)$ into the network. The process continues similarly for \textit{3)}, \textit{4)} and \textit{5)}. Similar to node $m$, nodes $o$ and $p$ will flood $(i^*,j^*,k^*,l^*,n^*,m,o^*)$ and $(i^*,j^*,k^*,l^*,n^*,m,o^*,p^*)$, respectively, after they determine the wake-up scheduling. Finally, every node has its wake-up scheduling and the location and position information of all other nodes in the self-rescue group. The location and position information will be included in the emergency message.

\algrenewcommand\algorithmicwhile{\textbf{when}}
\begin{algorithm}[!t]
\begin{footnotesize}
\caption{Self-rescue group communications protocol}
\label{alg:distributed}
\begin{algorithmic}[1]
\ForAll {$i \in V$}
\State broadcast \textit{hello} messages 
\State discover neighbors
\State \textcolor{red}{compute distances}
\State broadcast one-hop neighbors
\State construct one-hop network
\State find all maximal cliques and calculate $\theta_i$ and $\gamma_i$
\State flood $\theta_i$ and \textit{last known location} into network
\EndFor
\vspace{.15cm}
\If{$i=\arg\max_{j \in V}\theta_j$}
\State determine wake-up \textit{schedule}
\State \textcolor{red}{computing \textit{positions} of one-hop neighbors}
\State broadcast \textit{schedule} and \textcolor{red}{\textit{positions}}
\EndIf
\vspace{.15cm}
\ForAll{$i \in V$}
\While{received a \textit{schedule}}
\If{not already scheduled \&\& received enough schedules}
\If{received enough \textit{schedules}}
\State determine wake-up \textit{schedule}
\State \textcolor{red}{computing \textit{positions} of one-hop neighbors}
\If{wake-up scheduling not completed}
\State broadcast \textit{schedule} and \textcolor{red}{\textit{positions}}
\Else
\State \textcolor{red}{flood \textit{positions}}
\EndIf
\EndIf
\Else
\State continue to wait
\EndIf
\EndWhile
\EndFor
\end{algorithmic}
\end{footnotesize}
\end{algorithm}

The positioning works as an add-on function of self-rescue grouping and wake-up scheduling, and they are implemented together as the self-rescue group communications protocol as detailed in Protocol \ref{alg:distributed}. The procedures labeled in red are exclusively exploited for positioning, which are involved with computing positions and flooding position information. However, these procedures incur very little additional overhead. The computation overhead is low and positions are computed only once. Since only some nodes need to flood the position information, the communication overhead is also low. Moreover, the last known location is flooded together with $\theta$ and computed positions are broadcast with wake-up schedules. 

Note that all self-rescue nodes have to be awake during executing this communications protocol. However, every node knows when this process is complete, i.e., after it has received the position information of all other self-rescue nodes. Then, each node goes to sleep and wakes up accordingly based on the determined schedule.  

Self-rescue nodes have different battery levels and may consume different amounts of energy. However, each node in such a self-rescue group always consumes much less energy than if it stays awake alone, while maintaining a similar coverage. Since trapped survivors with a high battery level may or may not be willing to consume more energy than it has to when joining the self-rescue group, we currently do not consider how to balance the battery level among self-rescue nodes, but leave it for future work.

\subsubsection{Message Overhead}
\label{sec:overhead}
Let us analyze the message overhead of the self-rescue system according to Protocol \ref{alg:distributed}. Each node needs to broadcast three messages (\textit{hello} message, one-hop neighbors and wake-up scheduling) and flood a message of $\theta$. A node also needs to flood a message of position information if it is the last one in the maximal cliques it belongs to determine the wake-up scheduling. Let $N$ denote the number of self-rescue nodes. In the self-rescue group, there are $3N$ broadcast messages and flood messages are at most $2N^2-N$. Note that nodes that need to flood the position information are at most $N-1$. Therefore, the total message overhead of the self-rescue group $2(N^2+N)$, i.e., $2(N+1)$ transmissions per node. 

\subsubsection{Emergency Message}
Each self-rescue node acknowledges last known locations and position information of nodes in the group. This information will be included in emergency messages and sent to rescue crews. Therefore, an emergency message includes: \textit{(1)} the number of trapped survivors; \textit{(2)} the time when they were trapped; \textit{(3)} the latest known location of each node; \textit{(4)} the position of each node in the group.

\section{TeamPhone Implementation}
\label{sec:impl}

\begin{figure}[!t]
\setlength{\abovecaptionskip}{5pt}
\centering
\includegraphics[width=0.35\textwidth]{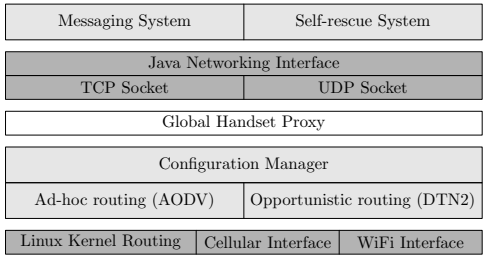}\\
\caption{Architecture of TeamPhone Implementation}
\label{fig:arch}
\end{figure}

In this section, we describe the detailed implementation, which includes TeamPhone interface, TeamPhone routing, and TeamPhone application, where the network interface configuration and routing are implemented in C, C++ based on Linux, and the application is implemented in Java based on Android. The architecture of TeamPhone implementation is illustrated in \figurename~\ref{fig:arch}. Note that the implementation requires the root privilege of the smartphone.  

\subsection{The Interface}

\begin{figure}[t]
\setlength{\abovecaptionskip}{3pt}
\centering
\includegraphics[width=0.35\textwidth]{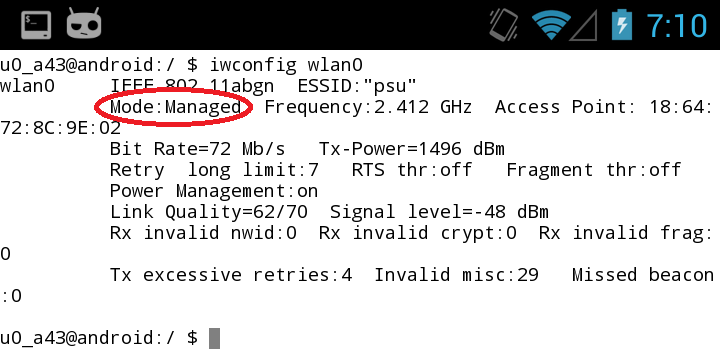} \\
\includegraphics[width=0.35\textwidth]{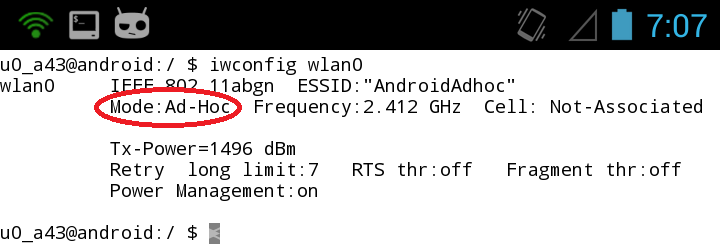}
\caption{WiFi status when being configure as managed mode and ad-hoc mode}
\label{fig:wifi}
\end{figure}

Besides the cellular interface, smartphones are usually equipped with Bluetooth and WiFi. Since Bluetooth has limited transmission range, WiFi is used for ad-hoc communications between smartphones. Although WiFi Direct or WiFi Tethering can be used to setup ad-hoc or opportunistic networking \cite{trifunovic2011wifi,raj2014darwin}, they are not feasible to support multihop communications, which is critical in disaster scenarios, and hence we choose WiFi ad-hoc mode. However, WiFi ad-hoc mode is not officially supported by Android. To enable WiFi ad-hoc mode, we need to compile wireless extension support into the Linux kernel and also compile the wireless tool \textit{iwconfig} for smartphones, which will be utilized to configure the WiFi driver. The configuration of WiFi includes switching WiFi from managed mode (also known as station infrastructure mode), which is the default mode of WiFi on off-the-shelf smartphones, to ad-hoc mode, and manipulating sleep and wake-up periods of WiFi. \figurename~\ref{fig:wifi} illustrates the status of WiFi on Samsung Galaxy S3 when being configured as managed mode and ad-hoc mode (the WiFi chip is BCM4330 from Broadcom). In TeamPhone, the ad-hoc routing exploits both the cellular interface (in gateway mode) and WiFi interface, while the opportunistic routing only employs the WiFi interface.

\subsection{TeamPhone Routing}

\textbf{AODV.}
In TeamPhone, we implemented AODV instead of proactive routing protocols (e.g., OLSR) and other on-demand source routing protocols (e.g., DSR) based on two major considerations: \emph{(i)} network topology changes are frequent in a disaster recovery scenario, which incurs high maintenance overhead including network traffic and power consumption for proactive routing; \emph{(ii)} on-demand source routing reacts slowly on path setup, restructuring and failures.

We modified the Linux implementation of AODV~\cite{aodvuu}, which also has some extra features (i.e., unidirectional link support and a signal quality threshold) to improve the performance of \emph{hello} messages, and cross-compiled it for smartphones. The implementation includes two components: a loadable kernel module and a user-space daemon. The kernel module captures data packets using \textit{Netfilter} with POSTROUTING, which requires the Linux kernel with \textit{Netfilter} enabled, and instructs the user-space daemon to issue a routing request if the destination of the packet is out of route entries. In addition, the kernel module is in charge of encapsulating and decapsulating the packet that goes through gateways. The user-space daemon maintains the kernel routing table and controls neighbor discovery, routing request and routing reply. The communications between these two components are based on \textit{NetLink} socket.

In disaster recovery, messaging nodes monitor the availability of cellular connections and configure themselves as gateways for AODV if the cellular connection is valid. To maximize the reachability of messages, if the destination cannot be reached by AODV, the message can be temporarily stored at the command center through gateways and the message will be delivered to the destination once it connects to the command center. We have integrated this function into the implementation of AODV. Table \ref{tab:parameter} shows the settings of major parameters of AODV in our implementation. 

\begin{table}[!b]
\setlength{\abovecaptionskip}{5pt}
\renewcommand{\arraystretch}{1.2}
\caption{Parameter settings of AODV}
\label{tab:parameter}
\centering
  \begin{tabular}{|c|c|}
    \hline
    \textbf{Parameter} & \textbf{Setting}  \\
    \hline\hline
    \textit{hello} broadcast interval & 1s \\
    \hline
    \# of \textit{hello}s before treating as neighbor & 3 \\
    \hline
    Allowed \textit{hello} loss & 2 \\
    \hline
    Routing request timeout & 4s  \\
    \hline
    Routing request retries & 2 \\
    \hline
    Routing reply ACK timeout & 50ms \\
    \hline
    Active route timeout & 10s \\
    \hline
  \end{tabular}
\end{table}

\textbf{DTN2.}
In TeamPhone, opportunistic routing is accomplished by DTN2~\cite{dtn2}, which is a Linux implementation of the DTN bundle protocol defined in RFC 5050~\cite{rfc5050}. We customized and cross-compiled DTN2 for smartphones. DTN2 is a platform that can adopt any opportunistic routing protocols, working at the application layer and running as a user-space daemon. The network interface configured for DTN2 is WiFi in ad-hoc mode. The basic workflow of DTN2 on smartphones is as follows. First, a convergence layer needs to be configured, which is used to transmit messages between smartphones. We choose the TCP convergence layer to guarantee the message transmission. Next, service discovery will advertise the convergence layer's presence to neighbors. Meanwhile a node also listens for neighbor beacons and distributes each event of neighbor discovery to the convergence layer. Then, DTN2 performs message transmissions with neighbors according to the configured routing protocol. Static routing and flood routing are two currently supported routing protocols in TeamPhone. 


\subsection{TeamPhone Application}
The TeamPhone application wraps the messaging system and self-rescue system together, implemented in Java on Android system. When TeamPhone is launched, WiFi will be configured as ad-hoc mode and users need to specify which system to use.

For the messaging system, AODV and DTN2 are initialized for routing, and the messaging app is provided for users to send and receive messages. Messages can be sent in three ways: \textit{(i)} through AODV to reach the destination; \textit{(ii)} by gateways to store the message at the command center from which the message will be eventually delivered when the destination connects to the command center (\emph{e.g.}, nodes periodically fetch messages from the command center directly or through gateways); \textit{(iii)} by DTN2 in static routing mode and flood routing mode. In the prototype, the first method is the default and other two are optional. The messaging system is also designed to receive emergency messages from self-rescue nodes and forward them to the command center. The \textit{hello} message from the messaging node and self-rescue node is flagged differently so as to deal with the unidirectional link between messaging node and self-rescue node, where only the messaging node can receive \textit{hello} messages from the self-rescue node. In that case, the messaging node will not receive the emergency message from the self-rescue nodes since the self-rescue nodes are unable to discover the messaging node. The messaging node can still be alerted by receiving the \textit{hello} message, and this alert function is implemented together with the user-space daemon of AODV.

The self-rescue system performs self-rescue grouping, wake-up scheduling and positioning in the background. For the case that trapped survivors may not have the opportunity to start self-rescue system manually, TeamPhone can employ \textit{iShake}~\cite{iShake}, which exploits smartphones as the seismic sensor\footnote{iShake needs to continuously measure the seismic wave. However, since a large earthquake typically lasts more than 30 seconds, it only needs to perform the measurement in a 30-second interval and hence its energy consumption should not be significant.}, to trigger the self-rescue system automatically when an earthquake occurs. After wake-up scheduling, self-rescue nodes can configure the determined wake-up schedule of WiFi using \textit{iwconfig}; \emph{i.e.}, they need to enable WiFi power management and set up the period between wake-ups and the timeout before going back to sleep (\emph{i.e.}, $\lambda$). When self-rescue nodes are awake, they run AODV to detect messaging nodes in their vicinity and send out the emergency message once they discover the messaging node. The function of sending emergency messages is embedded into the daemon of AODV. 

\section{Evaluations}
\label{sec:eval}

\begin{figure}[!t]
\setlength{\abovecaptionskip}{5pt}
\setlength{\belowcaptionskip}{-10pt}
	\centering
    \includegraphics[width=0.28\textwidth]{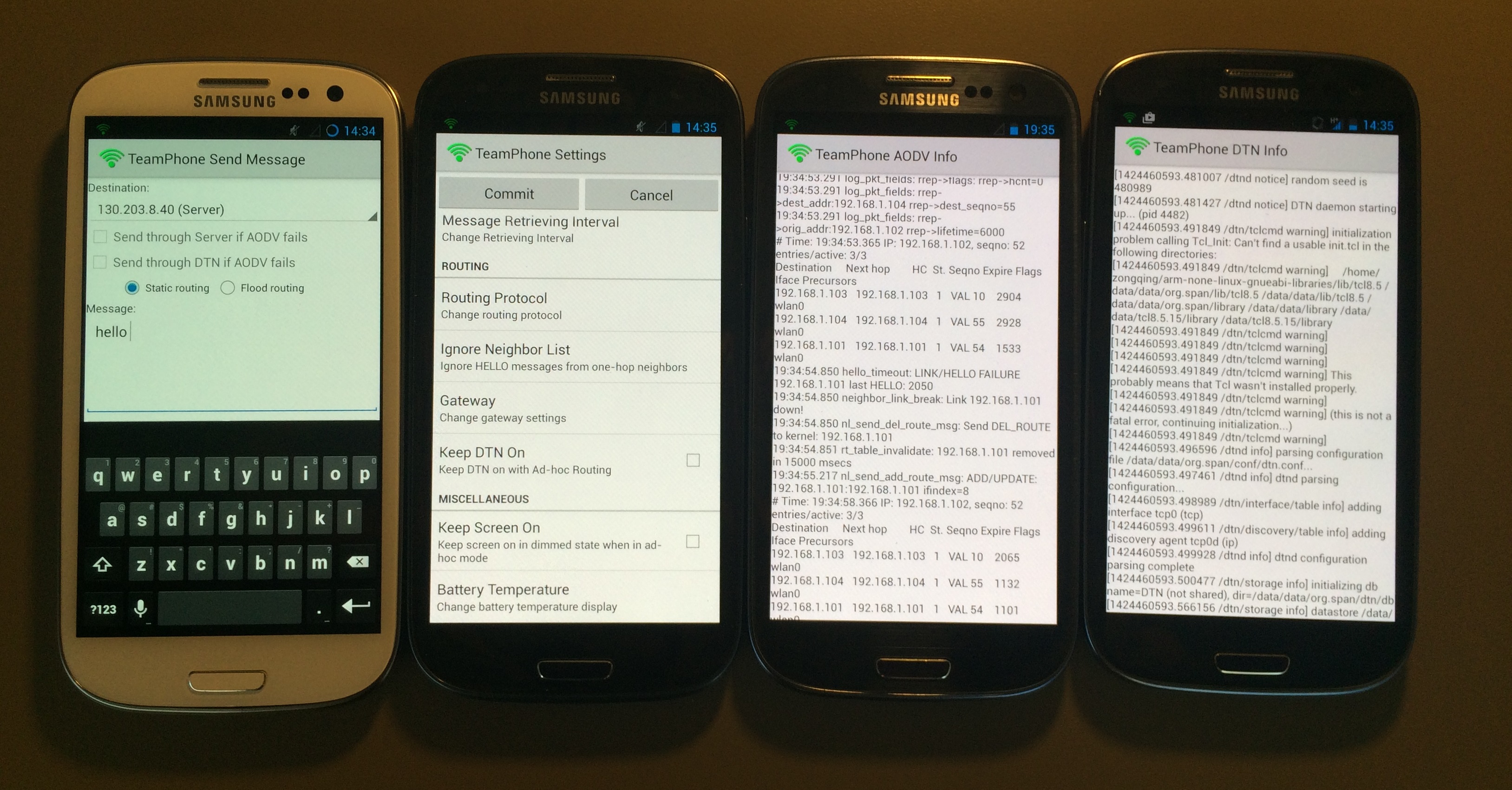}
    \caption{Testbed}
    \label{fig:testbed}
\end{figure}

TeamPhone is deployed on off-the-shelf Android smartphones, \emph{i.e.}, Samsung Galaxy S3. We built a small testbed of four S3 smartphones, as shown in \figurename~\ref{fig:testbed}, to evaluate the performance of TeamPhone. We do note that there exist designs for communications in disaster recovery. However, few are implemented as real systems and none provides system evaluation. Thus, we cannot compare them with TeamPhone in the evaluation. As discussed in Section \ref{sec:moti}, TeamPhone clearly outperforms these works in many aspects of design.

\begin{figure}[!t]
\setlength{\abovecaptionskip}{3pt}
\centering
	\begin{subfigure}{.155\textwidth}
		\centering
    	\includegraphics[width=1\textwidth]{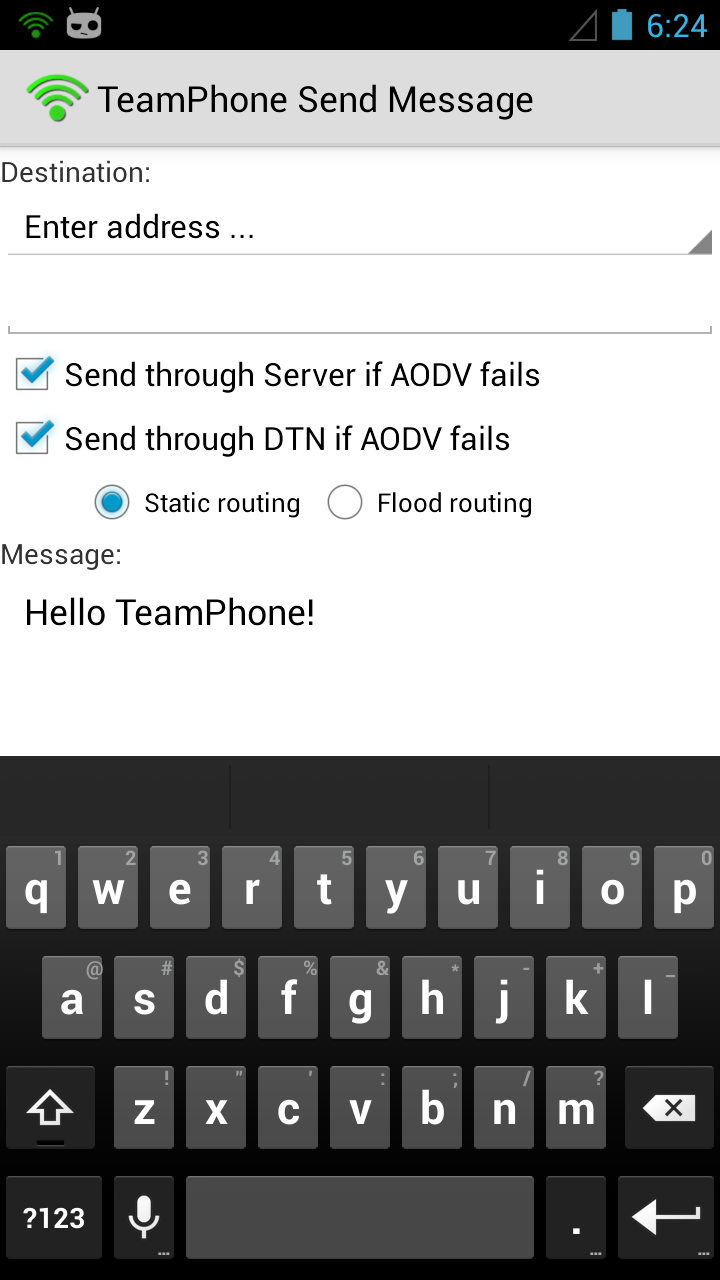}
    	\caption{sending message}
    	\label{fig:UI}
    \end{subfigure}
    \begin{subfigure}{0.16\textwidth}
		\centering    	
    	\includegraphics[width=0.96875\textwidth]{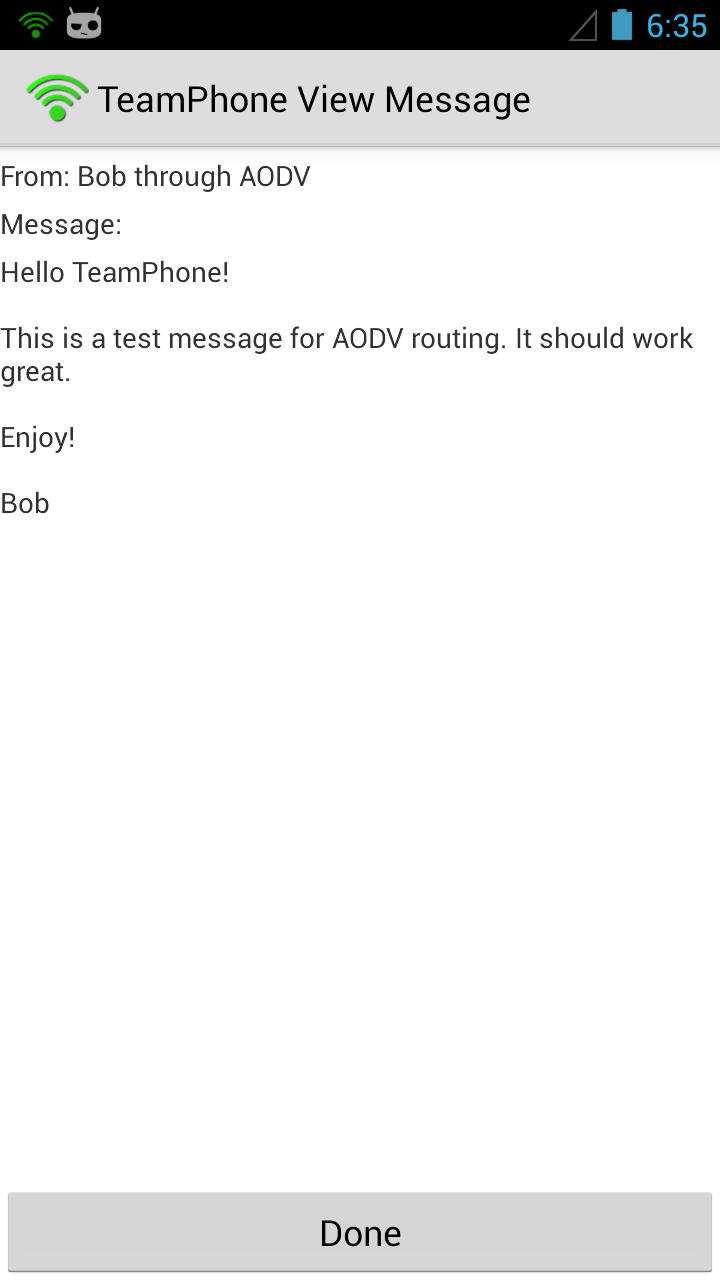}
    	\caption{viewing message}
    	\label{fig:recv}
    \end{subfigure}
    \begin{subfigure}{0.155\textwidth}
		\centering    	
    	\includegraphics[width=1\textwidth]{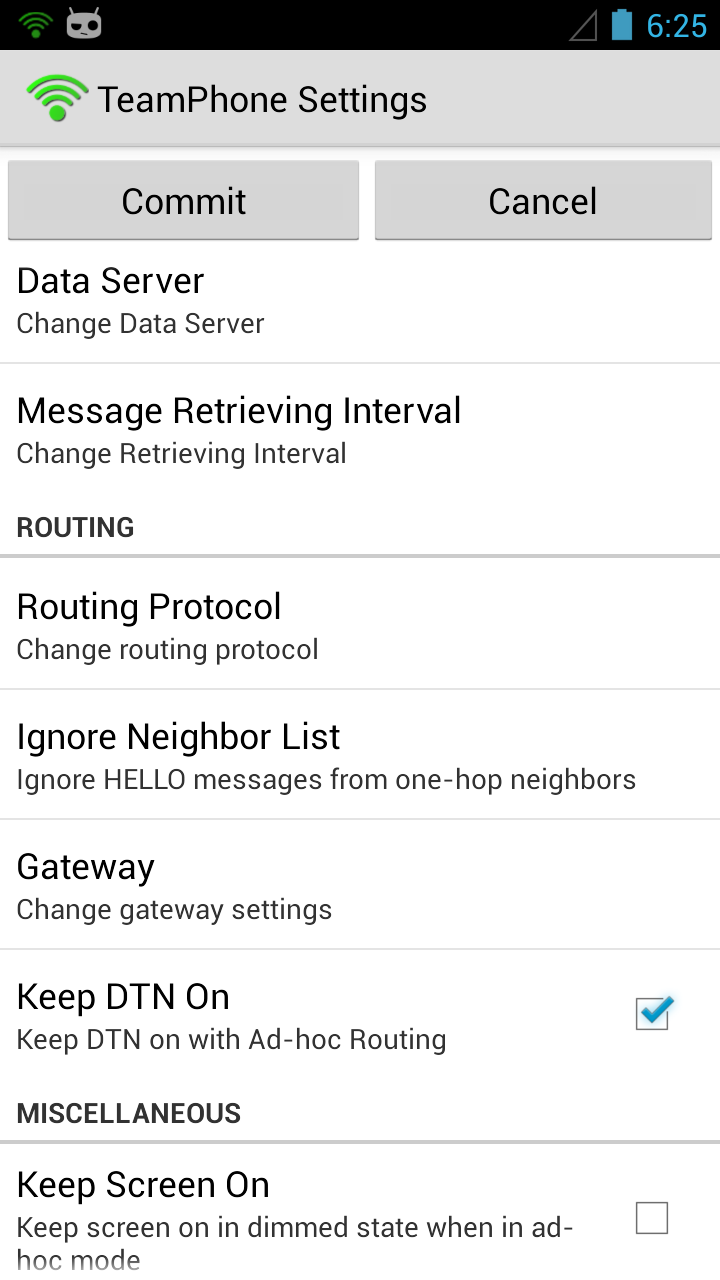}
    	\caption{settings}
    	\label{fig:settings}
    \end{subfigure}   
\caption{Messaging system}
\label{fig:msgsystem}
\end{figure}

\subsection{Messaging System}
\label{sec:messaging}
In the experiments, one smartphone has a cellular connection and works as a getaway. We also deploy a server which can be connected through the cellular network by smartphones to store and forward messages. This server acts as the command center in disaster recovery. \figurename~\ref{fig:msgsystem} illustrates the user interface for sending messages, viewing messages and settings\footnote{The messaging app is developed based on the open-source software MANET Manager \cite{manetmanager}.}. In \figurename~\ref{fig:UI}, the first checkbox corresponds to the ad-hoc routing with the getaway mode and the second checkbox corresponds to the opportunist routing. In the messaging system, messages are transmitted over TCP connections (TCP convergence layer for DTN2) to guarantee the delivery\footnote{TCP in ad hoc wireless networks may experience throughput degradation \cite{Murthy2004adhoc}, due to path length, packet loss, etc. However, since the messaging system is not throughout-intensive and requires more reliable transportation of packets, we choose TCP.}. The current implementation of TeamPhone can accomplish the following ways of message transmissions: \textit{i)} sending by AODV (including through gateways); \textit{ii)} sending by AODV-Gateway-Server-Gateway-AODV; \textit{iii)} sending by DTN2.
In the following, we evaluate the messaging system in terms of power consumption, throughput and delay, which are important aspects of a communication system in disaster recovery. We also show how opportunistic routing can facilitate messaging transmissions and obtain better performance than with only ad-hoc routing.

\begin{figure}[!t]
\setlength{\abovecaptionskip}{3pt}
\centering
\begin{minipage}[b]{.21\textwidth}
	\centering
    \includegraphics[width=0.88\textwidth]{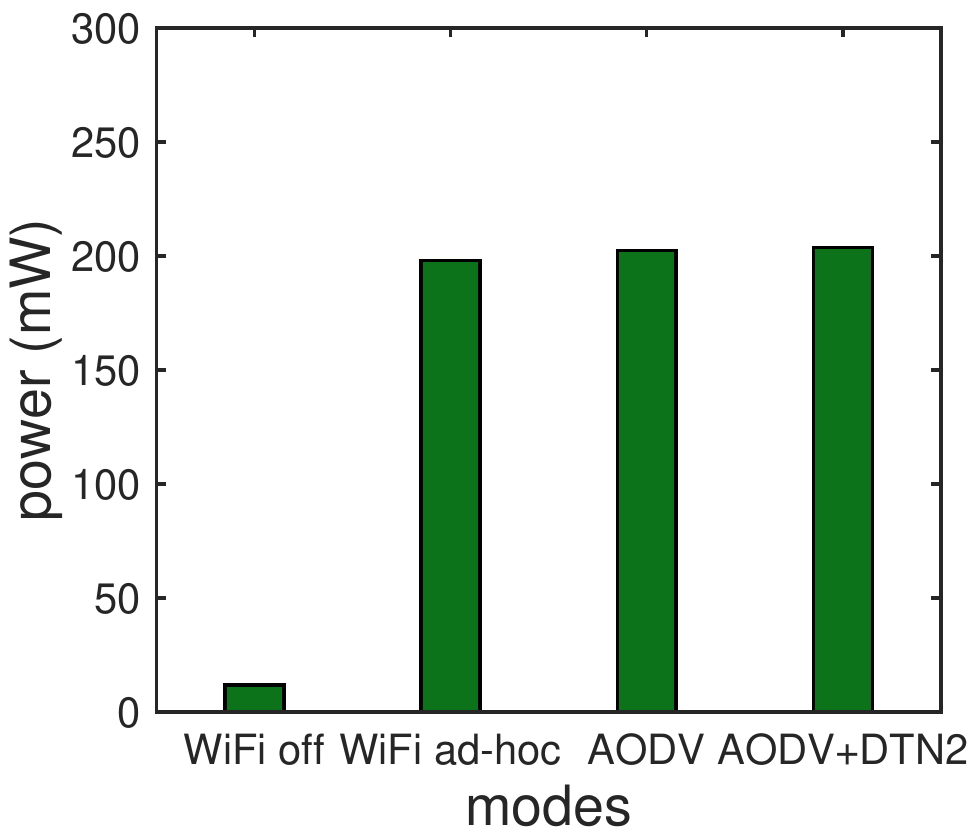}
    \caption{Power consumption when the smartphone is in different modes}
    \label{fig:power1}
\end{minipage}
\hspace{0.025cm}
\begin{minipage}[b]{.26\textwidth}
	\centering
    \includegraphics[width=0.7\textwidth]{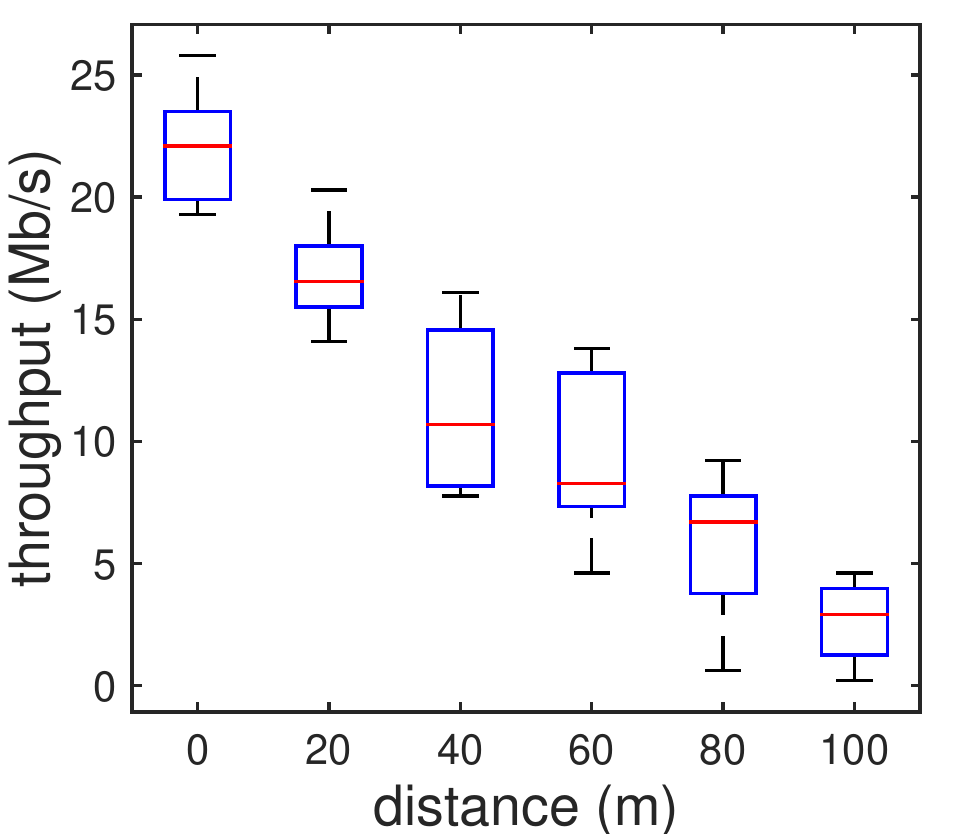}
    \caption{Throughput as a function of distance between two directly connected smartphones}
    \label{fig:distThroput}
\end{minipage}
\end{figure}

\textbf{Power.} The message system should be energy-efficient. To measure the power, the Monsoon Power Monitor is used to provide constant power to the smartphone instead of using battery.  \figurename~\ref{fig:power1} illustrates the power when the smartphone is operating in different modes, where all the measurements are conducted when the smartphone's screen is off and the power is averaged over one minute. As can be seen, when WiFi is off, the smartphone consumes about 10mW. When WiFi is on and configured in ad-hoc mode, the power consumption increases significantly, to about 200mW, because WiFi in ad-hoc mode has to be at the working stage to transmit and receive messages. When the messaging system with AODV (the \textit{hello} message interval is one second) is running on the smartphone without user-space data traffic, it incurs very little additional power. Similarly, the running of DTN2 only consumes very little power. From \figurename~\ref{fig:power1}, we can conclude that the basic power consumption of the messaging system (with both AODV and DTN2) is about 200 mW, which is mainly consumed by the WiFi module. For the Samsung Galaxy S3 with battery of 2100mAh and 3.8V, the estimated standby time is about 40 hours when running the messaging system.

\begin{figure}[!b]
\setlength{\abovecaptionskip}{3pt}
\centering
\begin{minipage}[b]{.21\textwidth}
	\centering
    \includegraphics[width=.85\textwidth]{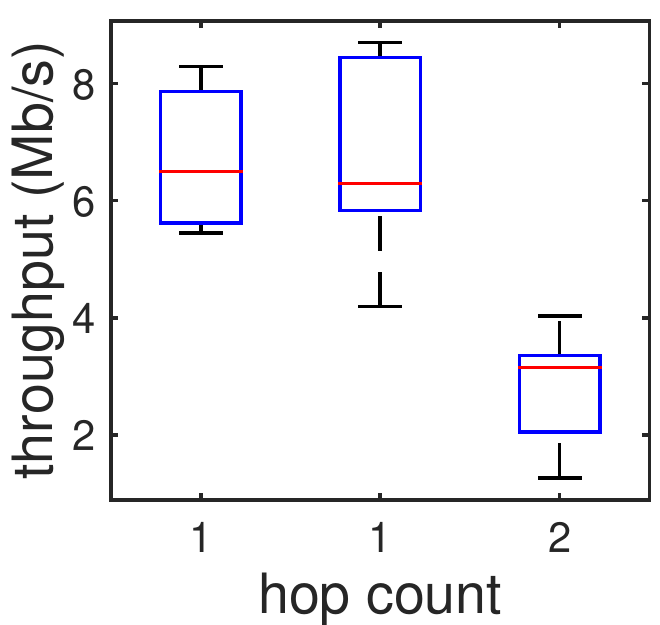}
    \caption{Throughput in terms of hop count}
    \label{fig:hopThroput}
\end{minipage}
\hspace{.2cm}
\begin{minipage}[b]{.21\textwidth}
	\centering
    \includegraphics[width=.85\textwidth]{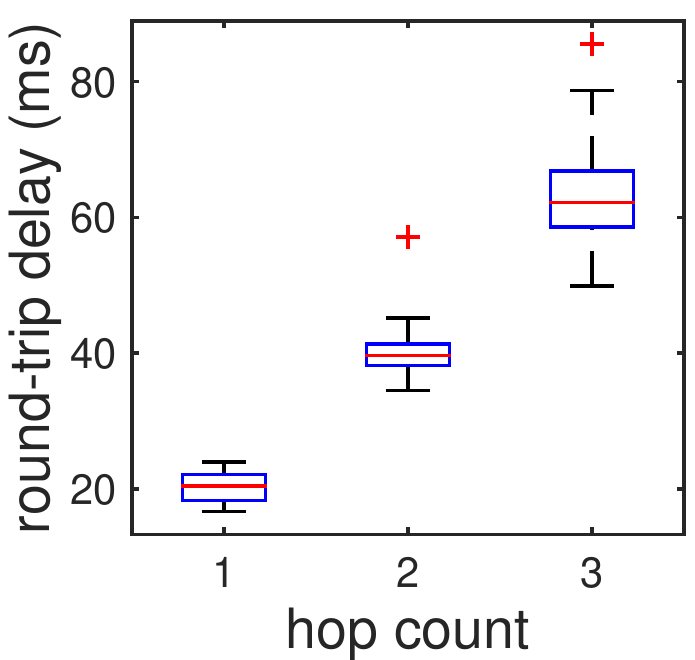}
    \caption{Round-trip delay of AODV on smartphones}
    \label{fig:delay}
\end{minipage}
\end{figure}

Table~\ref{tab:power2} gives the measured average power of the smartphone during one minute when AODV is configured with different \textit{hello} message intervals. As can be seen, the power for different intervals varies slightly. Although intuitively more frequent \textit{hello} messages should incur more power, the difference is too small compared to the power when WiFi is in ad-hoc mode (as shown in \figurename~\ref{fig:power1}). Therefore, the power is not the major concern when we choose the \textit{hello} message interval. 

\begin{table}[t]
\setlength{\abovecaptionskip}{5pt}
\renewcommand{\arraystretch}{1.2}
\caption{Power when AODV with different \textit{hello} message intervals}
\label{tab:power2}
\centering
\begin{tabular}{|c|c|c|c|c|c|}
\hline
\textit{\textbf{hello}} \textbf{interval} & 0.5s & 1s & 2s & 3s & 4s \\ \hline \hline
\textbf{Power (mW)} & 203.86 & 203.30 & 203.56 & 203.21 & 201.79  \\ \hline
\end{tabular}
\end{table}

The \textit{hello} message interval of AODV in TeamPhone is set to one second based on the following considerations. When AODV sends \textit{hello} messages more frequently, such as two \textit{hello} messages per second, it will cause more network traffic, especially when the network has many nodes. When \textit{hello} messages are sent less frequently, AODV will react to the change of network topology slowly and thus in turn affect the performance of AODV. For example, when the \textit{hello} interval is four seconds, the reaction delay to a neighbor change is about eight seconds, since the allowed loss of \textit{hello} messages is two as in Table \ref{tab:parameter}. Moreover, as observed from the experiments, the power consumption for different \textit{hello} message intervals only vary slightly. Therefore, in TeamPhone, the \textit{hello} message interval is set to one second.

\textbf{Throughput.} 
Since transmitted messages could be photo, voice, even video, we also measure the throughput of the messaging system 
as shown in Figures~\ref{fig:distThroput} and \ref{fig:hopThroput}, where the WiFi module of the smartphone S3 supports IEEE 802.11 a/b/g/n. 
\figurename~\ref{fig:distThroput} illustrates the throughput between two directly connected smartphones. The maximum throughout is over 20Mb/s, 
and it decreases linearly with the increase of the distance between two smartphones. When the transmission distance is 100 meters, 
the throughput drops to about 2Mb/s. \figurename~\ref{fig:hopThroput} shows the throughput on a two-hop ad-hoc path, constructed by AODV. 
As can be seen, the throughput of both one-hop paths is about 6Mb/s. 
The throughput of two-hop drops to 3Mb/s due to the increase of hop count.

\begin{figure}[!t]
\setlength{\abovecaptionskip}{3pt}
\centering
\begin{minipage}[t]{.2\textwidth}
	\centering
    \includegraphics[width=.8\textwidth]{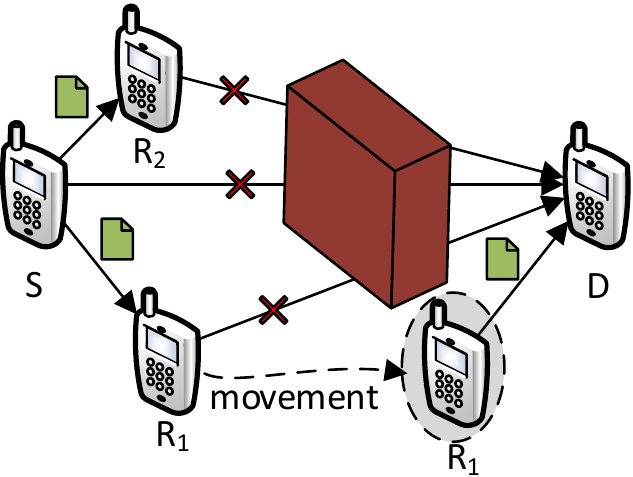}
    \caption{Movement helps data transmission}
    \label{fig:connectivity}
\end{minipage}
\hspace{.05cm}
\begin{minipage}[t]{.23\textwidth}
	\centering
    \includegraphics[width=1\textwidth]{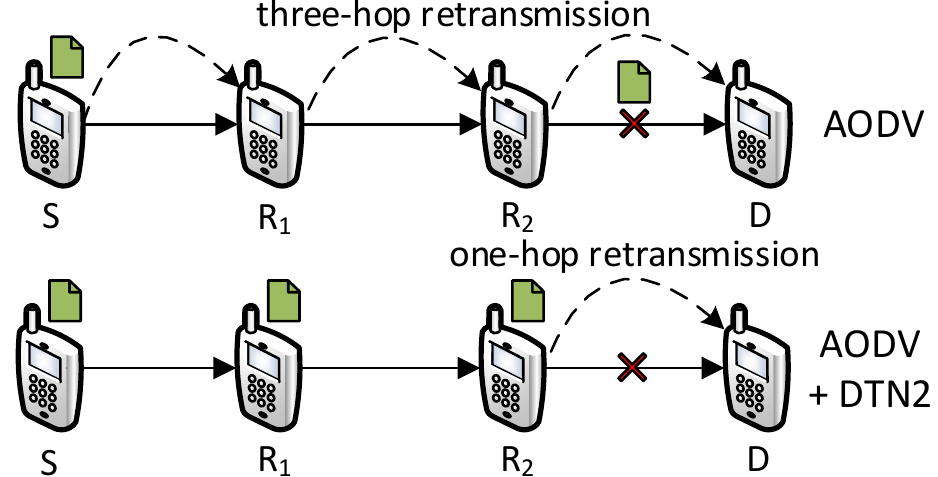}
    \caption{Replication reduces transmission delay}
    \label{fig:accleration}
\end{minipage}
\end{figure}

\textbf{Delay.} \figurename~\ref{fig:delay} shows the round-trip delay of message passing with AODV. As expected, the delay increases with the hop count, and the delay for one-hop message passing is low (about 20ms). For two-hop and three-hop, the delay is about 40ms and 60ms, respectively. Unlike one-hop, where neighbors are already in the routing table, multi-hop requires routing path discovery if the destination is not in routing table. Thus, the delay variations of two-hop and three-hop are much higher than the median values as shown \figurename~\ref{fig:delay}.

\textbf{Benefits of Opportunistic Routing.} The incorporation of opportunistic routing is designed to enhance the performance of the messaging system as demonstrated in \figurename \ref{fig:connectivity} and \ref{fig:accleration}.

\begin{figure*}[!t]
\setlength{\abovecaptionskip}{15pt}
\setlength{\belowcaptionskip}{-15pt}
\centering
	\begin{subfigure}[b]{.22\textwidth}
	\centering
    	\includegraphics[width=.9\textwidth]{TeamPhone/TeamPhone.11}
    	\caption{network topologies of self-rescue group}
    	\label{fig:topology}
    \end{subfigure}
    \hspace{.05cm}
    \begin{subfigure}[b]{0.21\textwidth}
	\centering    	
    	\includegraphics[width=.9\textwidth]{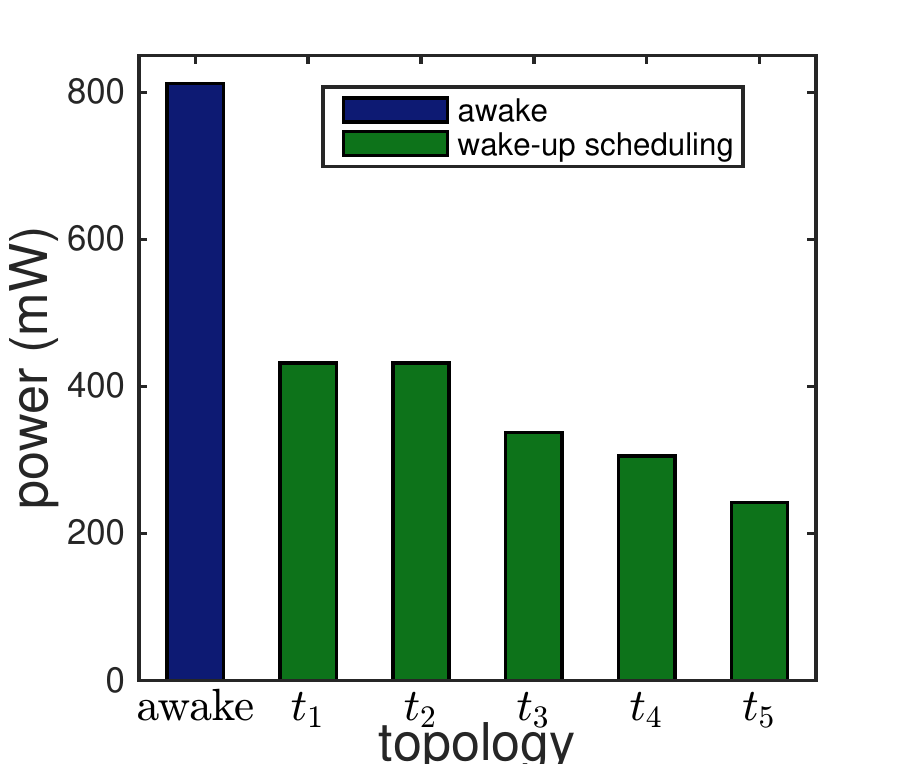}
    	\caption{total power of different topologies}
    	\label{fig:power4}
    \end{subfigure}
    \hspace{.05cm}
    \begin{subfigure}[b]{0.21\textwidth}
	\centering    	
    	\includegraphics[width=.9\textwidth]{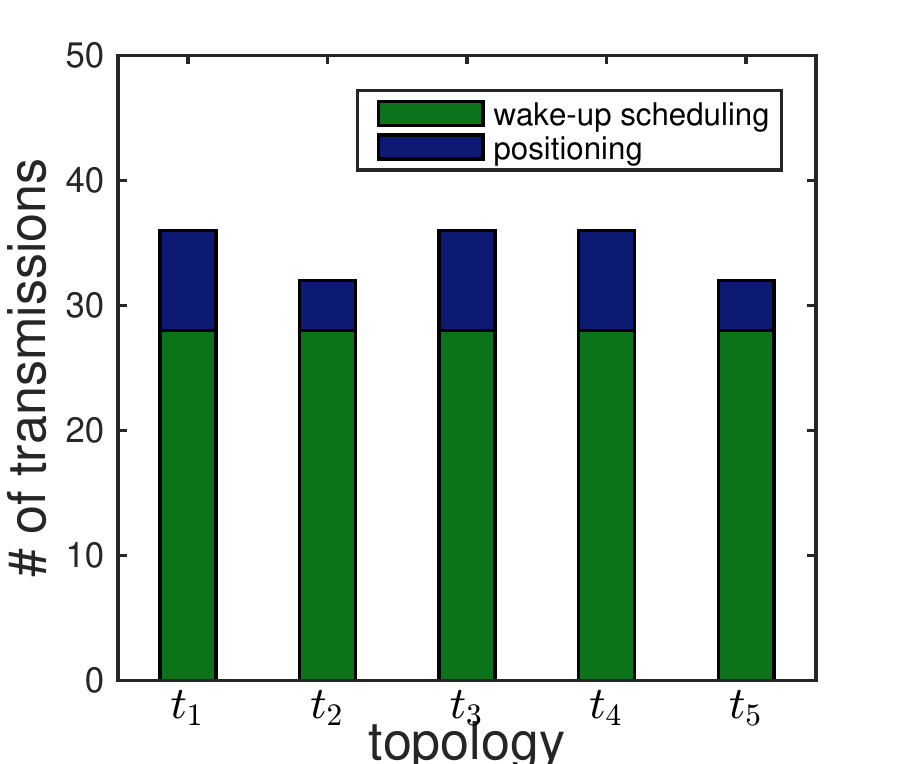}
    	\caption{message overhead of different topologies}
    	\label{fig:overhead-1}
    \end{subfigure}
    \hspace{.05cm}
    \begin{subfigure}[b]{0.21\textwidth}
	\centering    	
    	\includegraphics[width=.9\textwidth]{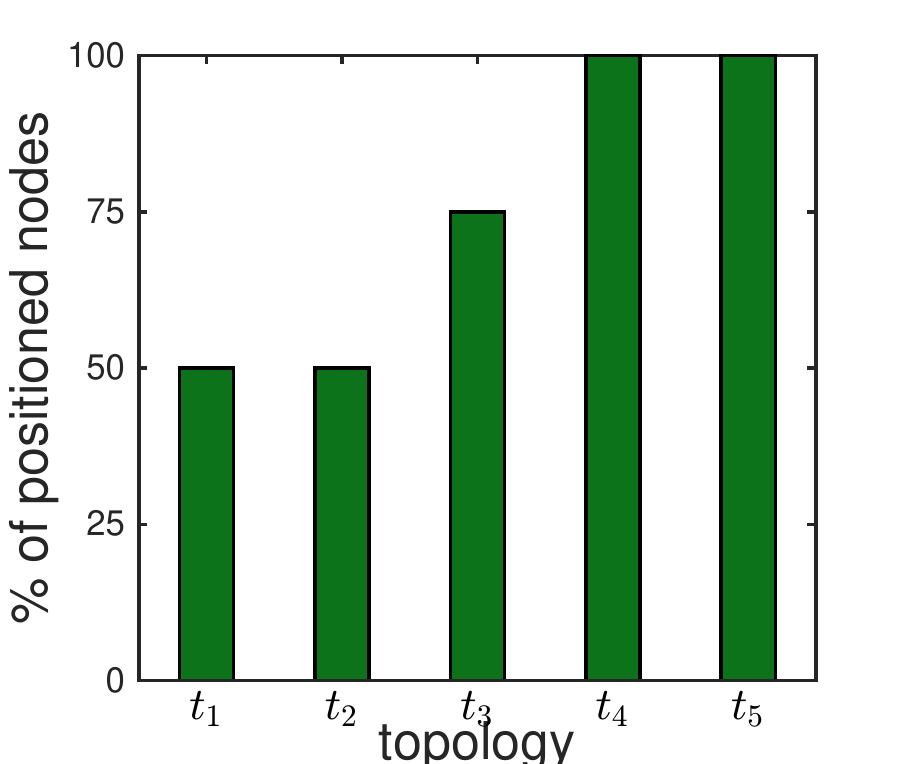}
    	\caption{percentage of positioned nodes in different topologies}
    	\label{fig:positioned}
    \end{subfigure}
\caption{Power, message overhead and positioned nodes of self-rescue groups in different topologies}
\label{fig:self-rescue-eval}
\end{figure*}

In \figurename \ref{fig:connectivity}, node $S$ needs to send a message to node $D$. Node $S$ first performs path discovery of AODV. However, neither $S$, $R_1$ nor $R_2$ can contact $D$ due to a physical obstacle as shown in \figurename \ref{fig:connectivity}. Therefore, the connection between $S$ and $D$ cannot be established using AODV (or any ad-hoc routing) and the message transmission is failed. Such a scenario is common in disaster recovery. Nevertheless, if opportunistic routing is deployed, the message can be replicated at $R_1$ and $R_2$ first since $R_1$ and $R_2$ can be reached from $S$, and then the message can be delivered when any of them moves closer to $D$ and establishes a connection with $D$. Therefore, opportunistic routing can take advantage of node movement to accomplish message transmissions; the delay is mainly determined by the physical movement.

In \figurename \ref{fig:accleration}, node $S$ is sending a message to $D$ through multi-hop connection between them. However, the link between $R_2$ and $D$ is suddenly broken due to channel interference, or physical obstacles. If only ad-hoc routing is used, $S$ has to re-transmit the message to $D$ when the link between $R_2$ and $D$ is back, and a new route discovery has to be initiated. With opportunistic routing, the messaging system can handle such a scenario with better performance as follows. When the link between $R_2$ and $D$ is broken, the messaging system can distribute the message to other nodes (\emph{i.e.}, $R_1$ and $R_2$) on the path to the destination. When the link is valid again, the message can be directly transmitted to node $D$ by $R_2$ instead of three-hop communications from $S$ to $D$. Therefore, message replication of opportunistic routing can greatly accelerate message transmissions. Based on the experiments on the testbed, for the scenario in \figurename \ref{fig:accleration}, the round-trip delay can be decreased from about 60ms to 20ms, and the throughput can be increased from about 1Mb/s to more than 6Mb/s.

From these two scenarios, we can see opportunistic routing can greatly facilitate message transmissions and improve the performance of the messaging system.

\subsection{Self-rescue System}
\label{sec:self}

For the self-rescue system, we first need to determine the wake-up period $\tau$. As self-rescue nodes wake up to discover the messaging node, $\tau$ should be long enough such that they can receive the \textit{hello} message from the messaging node and send out the emergency message. Since self-rescue nodes wake up alternately in a clique, $\tau$ should be short enough so that the self-rescue node is awake when messaging nodes are passing by. Based on these considerations and the parameter settings of AODV, $\tau$ is set to 5 seconds in the experiments.

\begin{table}[h]
\renewcommand{\arraystretch}{1.2}
\setlength{\abovecaptionskip}{5pt}
\caption{Power and energy consumption of the self-rescue system}
\label{tab:power3}
\centering
\begin{tabular}{|c|c|c|c|}
\hline
 & \textbf{Power (mW)} & \textbf{Energy (mJ)} & \textbf{Time period (s)} \\ \hline \hline
\textbf{Wake-up} & 202.30 & 1011.5 & 5 \\ \hline
\textbf{Sleep} & 12.98 & 64.9 & 5 \\ \hline
\end{tabular}

\end{table}  

\textbf{Power and Energy.} Table~\ref{tab:power3} shows the power and energy consumption of the self-rescue system. When nodes are asleep (WiFi goes to sleep), they consume very little power, similar to that when WiFi is turned off. When they are awake, the power consumption is similar to that of messaging nodes as in \figurename~\ref{fig:power1}. According to the wake-up scheduling, the worst case of energy consumption is when the sleep period of a self-rescue node is also $\tau$ (\emph{i.e.}, the self-rescue node only belongs to a two-node clique). In that case, the energy consumption is about 1076mJ for a sleep period and a wake-up period (totally 10 seconds) as shown in Table~\ref{tab:power3}.

To measure the power consumption of the self-rescue group, we use the four smartphones to form groups with different network topologies, as shown in Figure~\ref{fig:topology}. There are five different network topologies, denoted as $t_1$, $t_2$, $t_3$, $t_4$ and $t_5$. For each topology, we first determine the wake-up schedule and then measure the total power consumption of the group. In \figurename~\ref{fig:power4}, \emph{awake} indicates the power consumption when all the nodes stay awake. As can be seen in Fig~\ref{fig:power4}, our wake-up scheduling is always better than \emph{awake}, and the power consumption is less than half of \emph{awake}. When the network contains a larger clique, the self-rescue group consumes less power. For example, as $t_3$ contains a three-node clique while $t_2$ contains a two-node clique, the power consumption of $t_3$ is less than that of $t_2$. Similarly, $t_5$ (four-node clique) is less than $t_4$ (three-node clique). Moreover, the power consumption of $t_5$ is only $1/3$ of \emph{awake}. For smartphone S3, the standby time of the group when running the self-rescue system is about 70 hours for $t_1$ and 125 hours for $t_5$. Therefore, we can conclude that the self-rescue system can greatly save energy and thus increase the possibility of being discovered in rescue operations for trapped survivors.

\textbf{Message Overhead and Positioning.} \figurename~\ref{fig:overhead-1} shows the message overhead of the self-rescue system for different topologies. The message overhead includes both wake-up scheduling and positioning. As depicted in the figure, the overhead of wake-up scheduling is the same for different topologies since it only depends on the number of nodes in the self-rescue group. Meanwhile the overhead of positioning varies on different topologies, which depends on the number of nodes that belong to only one clique or are the last one in the cliques it belongs to determine the wake-up scheduling, as discussed in Section \ref{sec:overhead}. \figurename~\ref{fig:positioned} depicts the percentage of positioned nodes in these network topologies. For $t_1$ and $t_2$, half of nodes can be positioned; while all the nodes can be positioned in $t_4$ and $t_5$. Therefore, we can see that more nodes can be positioned if the group contains a larger clique.

\begin{figure}[!t]
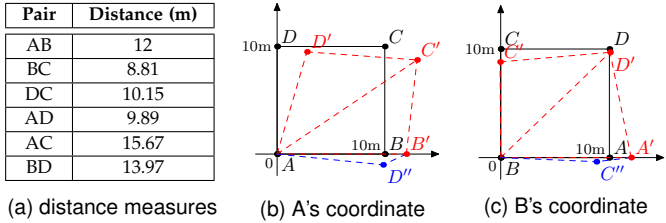

\setlength{\abovecaptionskip}{3pt}
\centering
    \begin{subfigure}{0.16\textwidth}
    	\centering
    	\begin{scriptsize}
		\renewcommand{\arraystretch}{1.1}
		\begin{tabular}{|c|c|}
		\hline
		\textbf{Pair} & \textbf{Distance (m)} \\ \hline\hline
		AB        & 12           \\ \hline
		BC        & 8.81         \\ \hline
		DC        & 10.15        \\ \hline
		AD        & 9.89         \\ \hline
		AC        & 15.67        \\ \hline
		BD        & 13.97        \\ \hline
		\end{tabular}
    	\end{scriptsize}
    	\caption{distance measures}
    	\label{fig:measured}
	\end{subfigure}
	\hspace{0.05cm}
	\begin{subfigure}{.15\textwidth}
	\centering
    	\includegraphics[width=.98\textwidth]{TeamPhone/position.0}
    	\caption{A's coordinate}
    	\label{fig:A-positioning}
    \end{subfigure}
    \hspace{0.05cm}
    \begin{subfigure}{0.15\textwidth}
	\centering    	
    	\includegraphics[width=.98\textwidth]{TeamPhone/position.1}
    	\caption{B's coordinate}
    	\label{fig:B-positioning}
    \end{subfigure}
\caption{Positioning in a self-rescue group (a four-node clique), where four smartphones form a 10m$\times$10m square.}
\label{fig:positioning}
\end{figure}

To investigate the performance of positioning in a self-rescue group, we use the four smartphones to form a 10m$\times$10m square (a four-node clique). The smartphones are placed as the black square in \figurename~\ref{fig:A-positioning}. Each node measures the distances to its neighbors using signal strength as discussed in Section~\ref{sec:position}. As there are two distance measures for each node pair, e.g., the measures of nodes $A$ and $D$ for $AD$, the smartphone will use the mean value as the distance. The distance measures on the smartphones are shown in \figurename~\ref{fig:measured}. \figurename~\ref{fig:A-positioning} shows the coordinate built by node $A$ using $AB'$ as the x-axis and the positions of other nodes ($B'$, $C'$ and $D'$). There is distortion of the positions due to the measurement error. $D''$ could be a possible position for $D$. However, as the measured distance between $BD$ (13.97m) is much larger than the distance of $B'D''$, $D''$ can be excluded. The edges in red are used by $A$ to calculate the position of each node, and the measured distance of $BD$ is used to exclude $D''$. \figurename~\ref{fig:B-positioning} shows the coordinate built by node $B$. Since the measure of $BD$ is more accurate than $AC$ (compare to $10\sqrt{2}$m), the distortion is alleviated. Similarly, $C''$ can be excluded as a possible position of $C$. As the node positions in the coordinates built by different nodes can be different, the self-rescue group has to use one coordinate to position all the nodes as discussed in Section~\ref{sec:position}.      

\begin{figure}[!t]
\setlength{\abovecaptionskip}{3pt}
\centering
	\begin{subfigure}{.24\textwidth}
	\centering
    	\includegraphics[width=.70\textwidth]{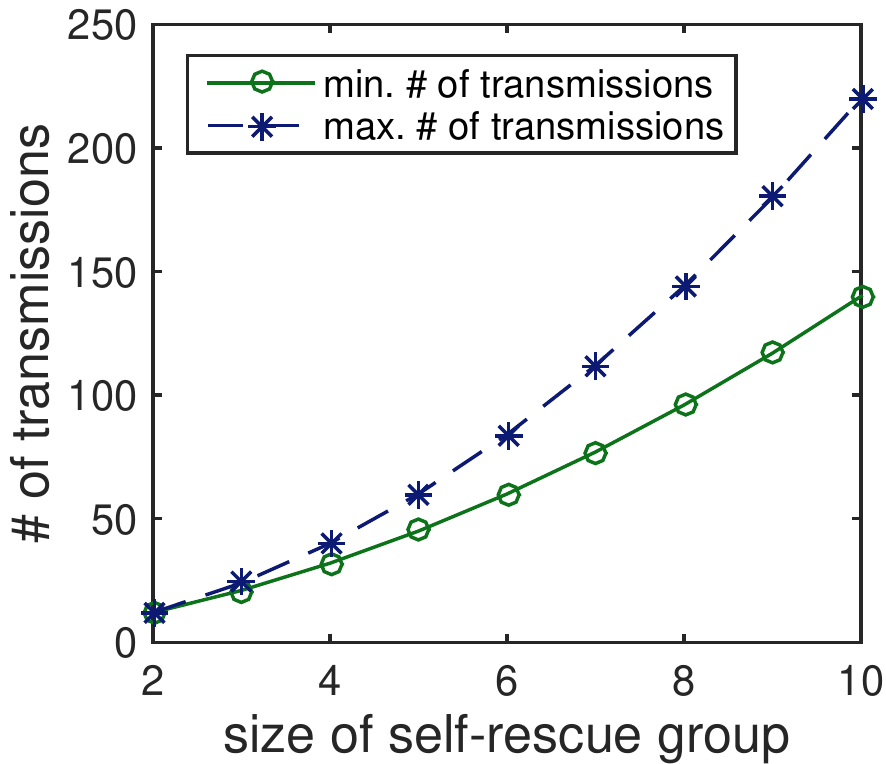}
    	\caption{total message overhead}
    	\label{fig:total}
    \end{subfigure}
    \begin{subfigure}{0.233\textwidth}
	\centering    	
    	\includegraphics[width=.70\textwidth]{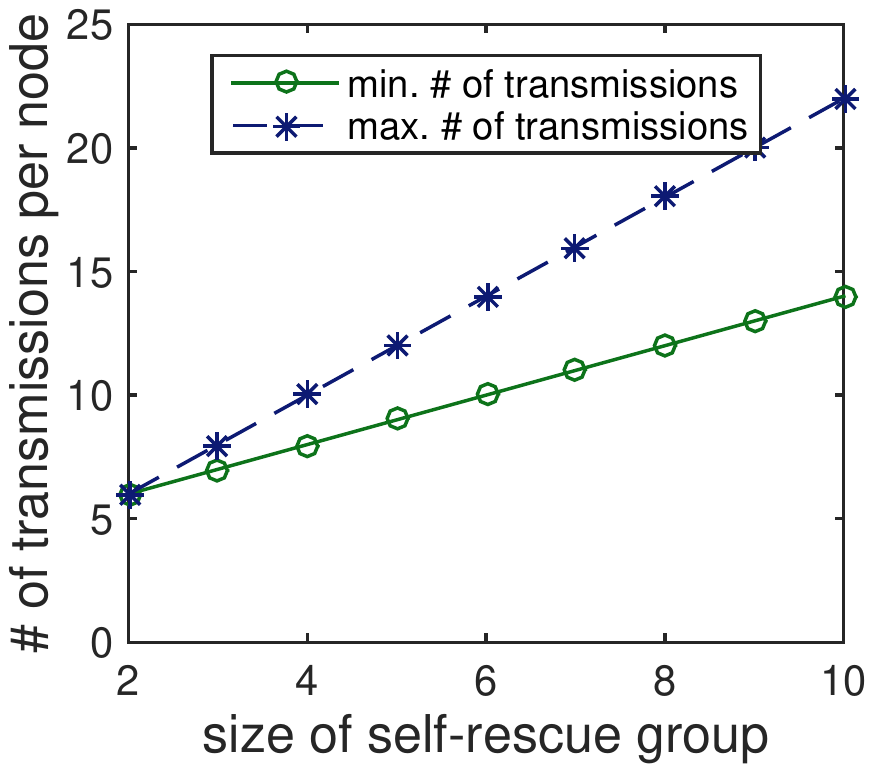}
    	\caption{message overhead per node}
    	\label{fig:pernode}
    \end{subfigure}
\caption{Message overhead in terms of the size of self-rescue group}
\label{fig:overhead}
\end{figure}

\figurename~\ref{fig:overhead} illustrates the message overhead of self-rescue system in terms of the size of self-rescue group. \figurename~\ref{fig:total} and \ref{fig:pernode} respectively give the total message overhead of the group and the message overhead per node. Since various topologies may have different message overhead, in the figure, we label the minimum and maximum overhead of all possible topologies for a certain size of self-rescue group. As the message overhead for a given network topology is fixed, we calculate the message overhead for each network topology based on the communications protocol for a certain size of self-rescue group. As depicted in \figurename~\ref{fig:total} and \ref{fig:pernode}, the message overhead increases with the size of the group. Although larger group incurs more message overhead, the overhead at each node just increases linearly and the self-rescue group is usually small in disaster recovery. With a few exchanged messages, self-rescue nodes are able to cooperatively and efficiently wake up and compose the emergency message that includes location and position information of the self-rescue group. 

\section{Related Work}
\label{sec:rela}

Research efforts have been made to apply communication technologies to disaster recovery. Many researchers focus on using ad-hoc networks. Zussman \textit{et al.} \cite{zussman2003energy} proposed to employ the network formed by smart badges to collect information from trapped survivors. Reina \textit{et al.} \cite{reina2011evaluation} extensively evaluated ad-hoc routing protocols in disaster scenarios. Aschenbruck \textit{et al.} \cite{aschenbruck2004human} modeled mobility in disaster areas and investigated how the model affected network performance of ad-hoc routing protocols. Other researches take opportunistic networks into consideration. Mart{\'i}n-Campillo \textit{et al.} \cite{Marti­nCampillo2013evaluating} proposed a random walk gossip protocol that runs over ad-hoc networks. Asplund \textit{et al.} \cite{asplund2009a} evaluated the efficiency of opportunistic routing protocols in disaster scenarios. Uddin \textit{et al.} \cite{uddin2013intercontact} proposed a multicopy opportunistic routing for disaster response networks. Fujihara and Miwa \cite{fujihara2014disaster} proposed disaster evacuation guidance using opportunistic communications. A more detailed survey of ad-hoc networks for disaster scenarios can be found \cite{reina2014survey}. 

Although many researchers have contributed to improving rescue and evacuation operations, only few consider smartphones. In the literature, \cite{suzuki2012soscast} and \cite{nishiyama2014relay} are the most related work to this paper. Suzuki \textit{et al.} \cite{suzuki2012soscast} proposed a design to assist the search for immobilized persons in a disaster area using Bluetooth of smartphones. However, Bluetooth of smartphones only has limited communication range (maximum 10m), and the design does not consider energy efficiency and can quickly drain the battery. 

Nishiyama \emph{et al.} \cite{nishiyama2014relay} assume that smartphones can be charged by mobile solar cells and thus did not consider energy efficiency. However, in disaster scenarios, solar cells may not be accessible for survivors, and thus the designed system can only provide temporary communications. Among the related works, few are implemented as real systems and none provides system evaluation. However, TeamPhone is designed, implemented, and evaluated based on off-the-shelf smartphones and ready to be installed on smartphones to provide communications and facilitate rescue operation in disaster scenarios.

\section{Conclusion}
\label{sec:conc}

In this paper, we propose TeamPhone, which is designed to network smartphones in disaster recovery. TeamPhone includes two components: the messaging system that provides data communications for rescue workers, and the self-rescue system that groups the smartphones of trapped survivors together, energy-efficiently discovers nearby messaging nodes and sends out emergency messages including location and position information. TeamPhone is implemented as a prototype application on the Android platform using the WiFi interface and cellular interface to provide several ways of communications. TeamPhone has been deployed and evaluated on the off-the-shelf smartphones. The evaluation results demonstrate that TeamPhone can accomplish various message transmissions with affordable power consumption and delay, and greatly reduce the energy consumption of sending out emergency messages by grouping and wake-up scheduling.


\ifCLASSOPTIONcompsoc
  \section*{Acknowledgments}
\else
  \section*{Acknowledgment}
\fi

This work was supported in part by Network Science CTA under grant W911NF-09-2-0053 and CERDEC via contract W911NF-09-D-0006. A preliminary version of this work appeared in the Proceedings of IEEE PerCom 2016 \cite{lu2016networking}.

\bibliographystyle{IEEEtran}
\bibliography{ref-long}

\end{document}